\documentclass[aps,superscriptaddress,eqsecnum,nofootinbib,preprintnumbers,longbibliography]{revtex4-2}
\usepackage{amsmath}
\usepackage{amssymb}
\usepackage{amsfonts,amsthm,bm}
\usepackage{color,xcolor}
\usepackage[greek,english]{babel}
\usepackage{titlesec}
\usepackage{graphicx,epsfig}
\usepackage{float}
\usepackage{subfigure}
\usepackage{tikz}
\usepackage{enumitem}
\usepackage{soul}
\usepackage[pdfencoding=auto,colorlinks=true,allcolors=blue]{hyperref}

\begin{document}
\title{Slowly-rotating compact objects: the nonintegrability of Hartle-Thorne particle geodesics}

\author{Kyriakos Destounis*}
\affiliation{Dipartimento di Fisica, Sapienza Università di Roma, Piazzale Aldo Moro 5, 00185, Roma, Italy}
\affiliation{INFN, Sezione di Roma, Piazzale Aldo Moro 2, 00185, Roma, Italy}
\email{kyriakos.destounis@uniroma1.it}
\author{and Kostas D. Kokkotas}
\affiliation{Theoretical Astrophysics, IAAT, University of T{\"u}bingen, 72076 T{\"u}bingen, Germany} 
\affiliation{Section of Astrophysics, Astronomy, and Mechanics, Department of Physics, National and Kapodistrian University of Athens, Panepistimiopolis Zografos GR15783, Athens, Greece}
	
\begin{abstract}
X-ray astronomy provides information regarding the electromagnetic emission of active galactic nuclei and X-ray binaries. These events provide details regarding the astrophysical environment of black holes and stars, and help us understand gamma-ray bursts. They produce estimates for the maximum mass of neutron stars and eventually will contribute to the discovery of their equation of state. Thus, it is crucial to study these configurations in order to enhance  the yield of X-ray astronomy when combined with multimessenger gravitational-wave astrophysics and black hole shadows. Nevertheless, an exact solution of the field equations does not exist for  rotating neutron stars. There exist a variety of approximate  solutions for compact objects that may characterize relativistic stars. The most studied approximation is the Hartle-Thorne metric that represents slowly-rotating compact objects, like massive stars, white dwarfs and neutron stars. Recent investigations of photon orbits and shadows of such metric revealed that it exhibits chaos close to resonances. Here, we thoroughly investigate particle orbits around the Hartle-Thorne spacetime up to second order in rotation. We perform an exhaustive analysis of bound motion, by varying all parameters involved in the system. We demonstrate that chaotic regions, known as Birkhoff islands, form around resonances, where the ratio of the radial and polar frequency of geodesics, known as the rotation number, is shared throughout the island. This leads to the formation of plateaus in rotation curves during the most prominent $2/3$ resonance, which confirms that generic geodesics are nonintegrable. We measure their width and show how each parameter affects it. The nonintegrability of Hartle-Thorne metric may affect quasiperiodic oscillations of low-mass X-ray binaries, when chaos is taken into account, and might potentially improve estimates of mass, angular momentum and multipole moments of astrophysical compact objects.
\end{abstract}
	
\maketitle

\section{Introduction}
X-ray observations have provided astrophysical explanations for various questions about compact objects in astrophysics, such as black holes (BHs) and neutron stars (NSs), particularly related to their surrounding astrophysical environments, including active galactic nuclei and low-mass X-ray binaries \cite{Taylor:1979zz,vanderKlis:2000ca,Kluzniak:2003ei,vanderKlis:2006}.

The motion of matter in the proximity of compact objects can be influenced by electromagnetic fields, which originate from a NS's boundary layer, an X-ray burst on the NS's polar caps, a hot corona around a BH, or an accretion disk in general (as described by the Poynting-Robertson effect; see, for example, \cite{Poynting:10.1093/mnras/64.1.1a,Robertson:10.1093/mnras/97.6.423}).
Observations from the Event Horizon Telescope that probe the environmental vicinity and shadow of M87* \cite{EventHorizonTelescope:2019dse} and Sgr A* \cite{EventHorizonTelescope:2022wkp} and gravitational-wave (GW) observations from the LIGO/Virgo/KAGRA collaboration \cite{LIGOScientific:2016aoc,LIGOScientific:2021djp} of BH-NS and NS binaries \cite{LIGOScientific:2017vwq,LIGOScientific:2021qlt} have paved the way for multi-messenger astrophysics. For binaries, this approach can combine both electromagnetic and gravitational knowledge \cite{Barack:2018yly}. By incorporating the experiments mentioned above, there are endless possibilities to explore new fundamental physics, gain a deeper understanding of the composition of astrophysical environments surrounding compact objects, and investigate the effects of dark matter halos in galaxies \cite{Barausse:2014tra,Cardoso:2021wlq,Cardoso:2022whc,Destounis:2022obl,Cheung:2021bol}. 

The binary NS merger GW170817 \cite{LIGOScientific:2017vwq}, was followed by a gamma-ray burst and a kilonova \cite{LIGOScientific:2017zic}. This observation was made through the use of electromagnetic telescopes across the spectra, and it confirmed that binary NSs are the progenitors of short gamma-ray bursts. Furthermore, it constrained the difference between the speed of light and the propagation speed of GWs, established new bounds on the violation of Lorentz invariance, and presented a new test for the equivalence principle.

In general, it is difficult to find any astrophysical object or system that is not spinning. As a result, they are not spherical and have specific multipolar structures. The degree of deviation from sphericity depends on their spin, environment, and other possible modifications to their composition. If the object is a spinning BH, then the spacetime is fully described by an exact solution of the Einstein field equations, the Kerr metric \cite{Kerr:1963ud}. In general, dealing with Kerr black holes is quite straightforward since they are only described by their mass and spin. On the other hand, compact rotating NSs are much more complicated and cannot be efficiently approximated by a Kerr exterior \cite{Berti:2003nb}. As of today, there is no analytical description of such spacetimes, and the geometry surrounding a rotating NS is much more complex because it directly depends on the highly unknown star's internal structure and its spin. Hence, the assumption that the exterior of a NS is similar to the Kerr metric is excessively simplistic and will likely lead to inaccurate conclusions regarding astrophysical processes occurring in their close proximity \cite{Pappas:2012nt}.

In order to achieve this goal, the scientific community has invested a significant effort in finding more precise approximations of NS exteriors, as evidenced by studies such as those referenced in \cite{Manko:1993CQGra..10.1383M,Manko:1995JMP....36.3063M,Manko:2000sg,Manko:2000ud,Stute:2002mqa,Pachon:2006an,Teichmuller:2011px,Pappas:2012nv,Pappas:2016sye}, with the eventual aim of using numerical methods to construct relativistic, highly accurate, and realistic NSs \cite{Stergioulas:1994ea,Stergioulas:2003yp}.  However, despite the advantages of numerical solutions of the field equations, such as increased reliability and adaptability to different symmetries and equation of state assumptions, analytic expressions remain preferable. The first and most extensively tested approximate solution to the field equations for the exterior of slowly-rotating, massive stars, white dwarfs or NSs, has been constructed perturbatively by Hartle and Thorne (HT) \cite{Hartle-Thorne2:1968,Hartle-Thorne1:1967,Hartle-Thorne3:1969,Hartle-Thorne4:1970,Hartle-Thorne5:1972,Hartle-Thorne6:1975}. This solution incorporates the Lense-Thirring effect \cite{Lense:1918,Vieira:2021ozg}, exhibits frame dragging,  leads to important precession effects \cite{Abramowicz:2003rc} and thus it has been specifically used to infer several astrophysical phenomena.

Observations of high-frequency flickers, known as quasiperiodic oscillations (QPOs), in the X-ray luminosity spectra of BHs and NSs, that result from the accretion process in low-mass X-ray binary systems, assist astronomers in comprehending the innermost regions of accretion disks \cite{vanderKlis:1985Natur.316..225V,Abramowicz:2002di,Pappas:2012nt,Ingram:2016tbq}. Additionally, they provide estimations of the masses, radii, and spin periods of white dwarfs, BHs, and NSs. These astrophysical investigations might include an accreting rotating NS, which can be described, for example, with the HT metric that engulfs matter from its donor star or accretion disk. This, in turn, aids in further understanding the frequency properties of radial and vertical epicyclic motion \cite{Urbancova:2019btk}, as well as geodesic motion \cite{Abramowicz:2003rc,Sulieva:2022kpc}. QPOs can be excited by various mechanisms and the current predominant models are (i) relativistic precession models \cite{Stella:2000gd}, where the QPO is directly related to the orbital motion and the radial or nodal precession of a particular geodesic, (ii) relativistic resonant models, where a resonance between the orbital and epicyclic frequencies is assumed and (iii) a variety of preferred radii models, where the geodesic of the fluid elements in the accretion disk is the source of the observed QPO frequency or the frequencies are produced from oscillatory modes of the entire accretion disk \cite{Rezzolla:2003zx}. One way or another, the aforementioned models, utilize properties of geodesics, thus the most relevant QPOs with respect to the orbital transient resonances that will be discussed in this work are the ones that connect to the orbital motion of the material in the accretion disk. Hence, a generic geodesic analysis around HT objects may as well connect and potentially affect the production of QPOs in low-mass X-ray binaries. However, potential chaotic phenomena resulting from the altered multipolar structure of HT objects have not been examined yet, neither their effects on geodesics of the accretion disk.

The Kerr metric, which is currently providing a good description of the shadow of M87* as measured by the Event Horizon Telescope, has integrable geodesics. This means that the motion of particles and photons around it can be separated into four first-order differential equations with the help of the Carter constant \cite{Carter:1968rr}. Integrability ensures the absence of chaos in null and timelike geodesics. Even though the Kerr solution is very special from many points of view, there are other Kerr-like solutions which also admit four constants of motion and thus their geodesics are integrable, even though the object's multipolar structure differs from that of Kerr (see e.g. \cite{Johannsen:2011dh,Papadopoulos:2018nvd,Konoplya:2018arm,Papadopoulos:2020kxu,Konoplya:2021slg}). However, there also exist compact objects that are non-Kerr in a sense that they do not possess four constants of motion and chaotic phenomena take place. This often occurs due to the absence of a Carter-like constant \cite{Glampedakis:2005cf} or a higher-rank Killing tensor. This may occur in alternative theories of gravity \cite{Cornish:2001jy,Cornish:2003ig,Verhaaren:2009md}, in non-vacuum BHs surrounded by astrophysical environments \cite{Barausse:2006vt}, in many body systems \cite{Barausse:2007dy}, in extreme-mass-ratio inspirals where the stellar-mass spinning secondary orbits a supermassive BH \cite{Suzuki:1996gm,Zelenka:2019nyp,Lukes-Gerakopoulos:2014dpa} and even when an external magnetic field is introduced in otherwise integrable BH spacetimes \cite{Stuchlik:2020rls,Stuchlik:2021gwg}; an attribute that is also present and extremely crucial in highly-magnetized NSs which lead to extraordinary phenomena such as fast radio bursts, gamma-ray bursts, and superluminous supernovae \cite{Leung:2022mvm}. As a result, a plethora of direct and indirect imprints of chaos occur. For instance, phase-space islands of trajectories appear around resonances (known as resonant islands) where the ratio of radial to polar frequency oscillations is shared throughout them, leading to a plateau in the frequency ratio evolution when a particle pierces a resonant island \cite{Apostolatos:2009vu,Lukes-Gerakopoulos:2010ipp,Destounis:2020kss,Chen:2022znf,Chen:2023gwm,Deich:2022vna}. Such indirect imprints of nonintegrability at the orbital level subsequently translate to abrupt jumps at the GW frequency evolution, known as GW glitches \cite{Destounis:2021mqv,Destounis:2021rko,Destounis:2023gpw,Destounis:2023khj} that may be detectable by future spaceborne GW detectors such as the Laser Interferometer Space Antenna (LISA) \cite{LISA:2017pwj,Barausse:2020rsu,LISA:2022yao,LISA:2022kgy,Karnesis:2022vdp}. Another recent example of indirect chaos for photon orbits is the realization of chaos in the shadows of highly deformed spacetimes, such as the HT metric, as discussed in \cite{Kostaros:2021usv}. These spacetimes form photon-trapping pockets that give rise to chaos, where photons are trapped near an orbital resonance between their radial and polar oscillations. This prolongs the trapping time and can produce chaotic fractal features in the corresponding shadows.

In this paper, we explore a significant portion of the parameter space of geodesics for massive particles around a HT compact object, up to second-order in rotation, denoted as $\mathcal{O}(\Omega^2)$ \cite{Hartle-Thorne1:1967,Hartle-Thorne2:1968} in order to spot potential transient resonances and resonant islands due to the possible chaoticity of HT compact objects. We do not restrict our study to circular or equatorial orbits, which have been examined in previous works in the HT metric \cite{Abramowicz:2003rc,Urbancova:2019btk,Sulieva:2022kpc} and quasi-Kerr perturbative solutions, where the perturbation to Kerr is the HT metric at $\mathcal{O}(\Omega^2)$ of the field equations \cite{Glampedakis:2005cf}. Instead, we investigate generic orbits in order to confirm the existence of chaotic phenomena associated to the nonintegrability of the HT spacetime. Our numerical findings confirm the existence of indirect chaos, such as the formation of Birkhoff chains in phase space \cite{Contopoulos_book}. Additionally, we observe islands of stability around periodic orbits and plateaus in the geodesic's radial ($\omega_r$) and polar ($\omega_\theta$) oscillation frequencies, which define the rotation curve. We performed a qualitative and quantitative analysis regarding the dependence of the widths of resonant islands of the $2/3$ resonance, by varying both the spacetime and particle parameters. In our study, we only focus on one of the most prominent resonances associated with extreme-mass-ratio inspirals \cite{Glampedakis:2005hs,Amaro-Seoane:2012lgq}, though we expect that other resonances exhibit similar chaotic phenomena as the ones presented here.

We observe that fixing all parameters and then varying the particle's energy (or $z$-component of its angular momentum) results in wider (or narrower) $2/3$ resonant islands and corresponding plateaus in the rotation curves. Additionally, we have found that increasing the spin and mass quadrupole deformation generally leads to wider islands, which agrees with previous research on non-Kerr metrics \cite{Glampedakis:2005cf,Apostolatos:2009vu,Lukes-Gerakopoulos:2010ipp,Lukes-Gerakopoulos:2012qpc,Lukes-Gerakopoulos:2014dpa,Destounis:2023gpw}, parameterized metrics \cite{Destounis:2020kss,Destounis:2021mqv,Destounis:2021rko}, and rotating boson stars \cite{Destounis:2023khj}. Accounting for these chaotic effects could provide further insight into the astrophysical environment of low-mass X-ray binaries, improve mass, spin, and multipole moment estimates of the BH or NS that consumes the donor star, and aid in the search for an ultimate equation of state for the supranuclear matter. We use geometric units in the following discussion where the gravitational constant $G$ and the speed of light $c$ are set to $G=c=1$.

\section{The Hartle-Thorne metric}

The HT metric describes stationary and axisymmetric spacetimes and can be used to approximate the interior and exterior of a rigidly, slowly-rotating compact object, such as a massive star, a white dwarf or a NS \cite{Hartle-Thorne1:1967,Hartle-Thorne2:1968}. It is not an exact solution of field equations but rather it is constructed perturbatively in terms of the rotation rate of the object. Specifically, even though we will operate only at the HT exterior approximation in $\mathcal{O}(\Omega^2)$, which is a perturbative solution to the field equation to second order in rotation, the particular metric has been carried out in higher order of spin, such as $\mathcal{O}(\Omega^3)$ \cite{Glampedakis:2017cgd,Kostaros:2021usv} and $\mathcal{O}(\Omega^4)$ \cite{Yagi:2014bxa}. This methodology is generically used to construct exotic compact object configurations \cite{Raposo:2018xkf,Pacilio:2020jza,Vaglio:2022flq,Vaglio:2023lrd} and perturbative solutions to modified theories of gravity \cite{Mignemi:1992nt,Mignemi:1993ce,Canizares:2012is,Deich:2022vna}. 

The perturbative form of the $\mathcal{O}(\Omega^2)$ HT metric in quasi-isotropic coordinates reads 
\begin{align}\label{metric}
	ds^2=-e^\nu \left(1+2h\right)dt^2+e^\lambda\left(1+\frac{2\mu}{r-2m}\right){dr^2}
	+r^2\left(1+2k\right)\{d\theta^2+\sin^2\theta\left[d\phi-\left(\Omega-\omega\right)dt\right]^2\}+\mathcal{O}(\Omega^3),
\end{align}
where we ignore higher than $\mathcal{O}(\Omega^2)$ terms in the rotation rate of the configuration, while the metric functions $h(r,\theta)$, $\mu(r,\theta)$, $k(r,\theta)$ and $\omega(r,\theta)$ can be expressed in terms of Legendre polynomials $P_\ell(\cos\theta)$ as
\begin{align}
	h(r,\theta)&=h_0(r)+h_2(r)P_2,\\
	\mu(r,\theta)&=\nu_0(r)+\mu_2(r)P_2,\\
	k(r,\theta)&=k_2(r)P_2,\\
	\omega(r,\theta)&=\omega_1(r)P^\prime_1.
\end{align}
By introducing $\chi=a/M=J/M^2$ as the dimensionless spin parameter with $a$ the spin, $J$ the angular momentum and $M$ the mass of the non-rotating object, we can significantly simplify the metric tensor components of spacetime. Furthermore, we can define a quadrupole deviation $\delta q$ from the Kerr quadrupole that leads to $Q=-\chi^2 M^3 (1-\delta q)$ \cite{Hartle-Thorne1:1967,Hartle-Thorne2:1968,DeFalco:2020vli,DeFalco:2021nfk,Kostaros:2021usv}\footnote{In higher order perturbative solutions, further deformation parameters are introduced from the perturbative field equations, see e.g. the discussion in Section 6 of \cite{Kostaros:2021usv}, where, besides the mass quadrupole, the spin-octupole parameter of the $\mathcal{O}(\Omega^3)$ HT metric also deviates from Kerr's spin octupole.}. The term $\left(1-\delta q\right)$  means that the larger $\delta q$ is, the more prolate the object becomes along the direction of the spin axis, leading to a more positive quadrupole than that of Kerr. Prolate deformations are what one expects from an astrophysical object so as to produce ultra-compact BH mimickers, according to \cite{Glampedakis:2017cgd}. Moreover, HT spacetimes with positive $\delta q$ have been found to exhibit interesting features, such as off-equatorial photon spheres and triple photon spheres that in specific regions of the available parameter space lead to stable trapping and subsequent chaotic behavior in the HT spacetime in study as well as other configurations such as boson stars and scalarized BHs \cite{Cunha:2016bjh,Shipley:2016omi}. Therefore, we will use as a guide the aforesaid analyses regarding photons and focus on the maximization of choatic imprints by choosing $\delta q >0$ for our massive test-particle analysis, thus the HT configurations we will be dealing with in the rest of this work will be oblate. In the special case when $\delta q=1$, $Q=0$. Nevertheless, the object is not spherical due to its non-zero rotation $\chi\Rightarrow g_{t\phi}\neq 0$. Stated differently, the metric indicates that there is still a prolate morphology because of the presence of a non-zero (positive) quadrupole deformation. The object's spin, together with the prolateness introduced by $\delta q=1$, counterbalance each other in the mass quadrupole to give $Q=0$, but the object is still spinning, which is a typical behavior of prolate rotating compact objects. 

We define the second-order correction to the mass $\delta m$ due to the deformation and spin, and a dimensional radial coordinate $x=r/M$ to obtain the HT metric as follows \cite{Benhar:2005gi,Glampedakis:2018blj,Kostaros:2021usv}:
\begin{align*}
	m&=M,\,\,\,\,\,\, e^\nu=e^{-\lambda}=1-\frac{2}{x}, \,\,\,\,\,\, \omega_1=\Omega-\frac{2\chi}{M x^3},
	\mu_0=M \chi^2\left(\delta m-x^{-3}\right), \,\,\,\,\,\,h_0=\frac{\chi^2}{x-2}\left(x^{-3}-\delta m\right),\\
	h_2&=\frac{5\chi^2}{16}\delta q\left(1-\frac{2}{x}\right)\left[3x^2\log\left(1-2/x\right)
	+\frac{(2-2/x)}{x(1-2/x)^2}(3x^2-6x-2)+\frac{\chi^2}{x^3}\left(1+\frac{1}{x}\right)\right],\\
	k_2&=-\frac{\chi^2}{x^3}\left(1+\frac{1}{x}\right)-\frac{5\chi^2}{8}\delta q
	\left[3\left(1+x-\frac{2}{x}-3\left(1-\frac{x^2}{2}\right)\right)\log\left(1-\frac{2}{x}\right)\right],\\
	\mu_2&=-\frac{5M x \chi^2}{16}\delta q \left(1-\frac{2}{x}\right)^2\left[3x^2\log\left(1-\frac{2}{x}\right)
	+\frac{(2-2/x)}{x(1-2/x)^2}(3x^2-6x-2)-\frac{\chi^2}{x^2}\left(1-\frac{7}{x}+\frac{10}{x^2}\right)\right].
\end{align*}
For simplicity, we assume that the total mass of spacetime is $M\equiv m+\delta m$ that includes the correction due to $\delta q$ and the small spin $\chi=a/M$ and basically amounts to setting $\delta m=0$ in the aforementioned equations. Lastly, the composition of matter that constructs the HT compact object is assumed to be such that usual issues with photon emission are evaded, so that the metric can be considered as a BH mimicker \cite{Cardoso:2019rvt,Maggio2020}. 

The metric described by \eqref{metric} reduces to the approximate Kerr spacetime, up to $\Omega^2$, when $\delta q=0$ and an appropriate transformation is imposed \cite{DeFalco:2021nfk,Glampedakis:2018blj,Bini:2013tia}. It also possesses an ergosphere when $g_{tt}=0$ \cite{Abramowicz:2003rc} while when the spin is set to zero, it reduces to Schwarzschild, since $Q=0$, $\omega_1=\Omega$ and $h=\mu=k=\omega=0$ in \eqref{metric}. We will assume that with respect to calculating geodesics (presented below), the spacetime metric is ``as it is given'' in Eq. \eqref{metric}, that is, the spacetime is described by the truncated to the second order metric and no further approximations are made. For more details regarding regions with potential pathological features of the HT metric (which we avoid in this work) see \cite{Kostaros:2021usv}.

\section{Geodesics and chaotic imprints}

The motion of test-particles on a general, stationary and axisymmetric spacetime of the form
\begin{equation}
	\label{line_element}	ds^2=g_{tt}dt^2+2g_{t\phi}dtd\phi+g_{rr}dr^2+g_{\theta\theta}d\theta^2+g_{\phi\phi} d\phi^2,
\end{equation}
is described by the geodesic equations
\begin{equation}\label{geodesic_equation}
	\ddot{x}^\kappa+\Gamma^\kappa_{\lambda\nu}\dot{x}^\lambda\dot{x}^\nu=0,
\end{equation}
where $\Gamma^\kappa_{\lambda\nu}$ are the Christoffel symbols, $x^\kappa$ is the four-position of the timelike orbit, and the overdots denote differentiation with respect to proper time $\tau$\footnote{Even though we will not be dealing with GW emission in this work, one can tranform from proper to coordinate time $t$ with a simple chain rule such that $r^\prime=dr/dt=(dr/d\tau)(d\tau/dt)=\dot{r}/\dot{t}$.}. From stationarity and axisymmetry, the spacetime axiomatically possesses two Killing vector fields, namely the conserved energy $E$ and $z$-component of the angular momentum $L_z$, described through Eq. \eqref{geodesic_equation}, as
\begin{equation}\label{energy_momentum}
	-E/\mu=g_{tt}\dot{t}+g_{t\phi}\dot{\phi},\,\,\,\,\,\,\,\,\,
	L_z/\mu=g_{t\phi}\dot{t}+g_{\phi\phi}\dot{\phi},
\end{equation}
where $\mu$ is the conserved mass of the test particle (not to be confused with the functions of spacetime $\mu(r,\theta)$). From Eq.~\eqref{energy_momentum}, we obtain two first-order differential equations for the $t$- and $\phi$-momenta as
\begin{equation}\label{tphi_dot}
	\dot{t}=\frac{E g_{\phi\phi}+L_zg_{t\phi}}{\mu\left(g^2_{t\phi}-g_{tt}g_{\phi\phi}\right)},\,\,\,\,\,\,\,\,\,
	\dot{\phi}=\frac{E g_{t\phi}+L_zg_{tt}}{\mu\left(g_{tt}g_{\phi\phi}-g^2_{t\phi}\right)}.
\end{equation}
Test particles in geodesic motion conserve, besides $E$ and $L_z$, their mass (or equivalently their four-velocity $g_{\lambda\nu}\dot{x}^\lambda\dot{x}^\nu=-1$), which leads to a constraint equation that defines bound motion:
\begin{equation}\label{constraint_equation}
	\dot{r}^2+\frac{g_{\theta\theta}}{g_{rr}}\dot{\theta}^2+V_\text{eff}=0,
\end{equation}
with $V_\text{eff}$ being a Newtonian-like effective potential
\begin{equation}
	V_\text{eff}\equiv\frac{1}{g_{rr}}\left(1+\frac{g_{\phi\phi} E^2+g_{tt}L_z^2+2g_{t\phi}E L_z}{\mu^2\left(g_{tt}g_{\phi\phi}-g_{t\phi}^2\right)}\right).
\end{equation}
When $V_\text{eff}=0$ the resulting curve is called curve of zero velocity~(CZV) since $\dot{r}=\dot{\theta}=0$ need to be satisfied. In general, the radial and polar motion is described by second-order coupled differential equations which are solved simultaneously. In the special case where there exists a fourth constant of motion, such as the Carter constant in Kerr spacetime, the system is characterized as integrable. All four equations for the motion of $(t,r,\theta,\phi)$ are described by first-order, decoupled equations, rendering the system deterministic. However, integrability is a fragile property and can be easily broken.

\begin{figure*}[t]\centering
	\includegraphics[scale=0.35]{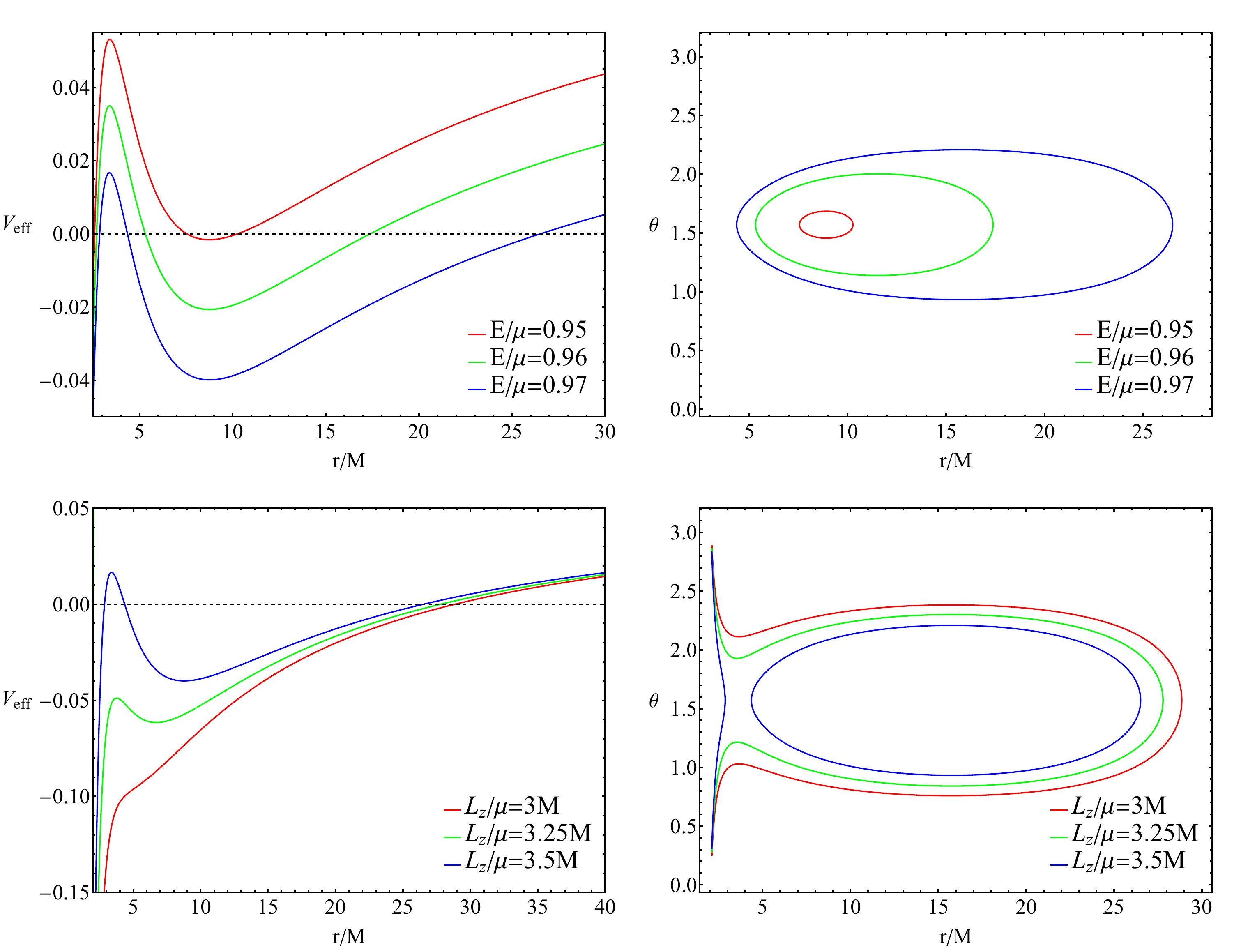}
	\caption{Top row: Equatorial effective potential $V_\text{eff}$ and corresponding CZV of the HT metric for varying $E/\mu$ and fixed $L_z/\mu=3.5M$, $\chi=0.3$, $\delta q=1$. Bottom row: Equatorial effective potential $V_\text{eff}$ and corresponding CZV for varying $L_z/\mu$ and fixed $E/\mu=0.97$, $\chi=0.3$, $\delta q=1$.}
	\label{fig:potential_CZV}
\end{figure*}

Nonintegrability implies that there might be mild or strong chaotic phenomena in the system's evolution depending on its parameters \cite{Contopoulos:2011dz,Contopoulos_book}. In these cases, the Kolmogorov-Arnold-Moser (KAM) theorem \cite{Arnold_1963,Moser:430015} states that \emph{if the orbits lie sufficiently away \cite{Contopoulos_book} from periodic points then its evolution is slightly shifted from that of the corresponding integrable system but presents no chaotic behavior}. For these orbits, curves that define successive intersections on a surface of section called a Poincar\'e map \cite{Contopoulos_book}, namely the KAM curves, form closed curves and surround a periodic central point of the Poincar\'e map.  \emph{On the other hand, the Poincar\'e-Birkhoff theorem \cite{Birkhoff:1913} dictates that the structure of a resonant KAM curve is disintegrated into an even number of stable and unstable periodic points that form a Birkhoff chain when the orbit is periodic}. Around stable periodic points, a series of resonant KAM curves form, defining Birkhoff islands of stability (resonant islands). These KAM curves do not surround the central point of the map but rather the stable periodic points of the resonance. The unstable periodic orbits give rise to chaotic orbits that shield the stability islands with thin chaotic layers. The structure, as a whole, discussed above for resonant orbits of nonintegrable systems is called a Birkhoff chain. 

The most important aspect of resonant islands is the fact that particle orbits that reside inside them share the same periodicity with the central periodic orbit of the island, namely the ratio of the radial over the polar frequencies $\omega_r/\omega_\theta$, or the rotation number. Integrable systems do not exhibit Birkhoff chains, but still have resonant points. The difference between a resonant island of a nonintegrable system and a resonant orbit of an integrable system lies to the fact that the integrable resonance occupies a zero-volume of phase space while the non-integrable resonant island induces a finite volume of phase space where the rotation number is shared for all geodesics inside the island and the trajectory is locked in perfect resonance. This gives rise to an indirect imprint of chaos (due to the underlying but mostly indistinguishable chaotic layers, though see \cite{Destounis:2023khj}) and have been found to affect the orbital evolution \cite{Apostolatos:2009vu,Lukes-Gerakopoulos:2010ipp} and GW emission \cite{Destounis:2020kss,Destounis:2021mqv,Destounis:2021rko,Destounis:2023gpw,Destounis:2023khj} from test-particles orbiting around nonintegrable spacetimes.

\begin{figure*}[t]\centering
	\includegraphics[scale=0.31]{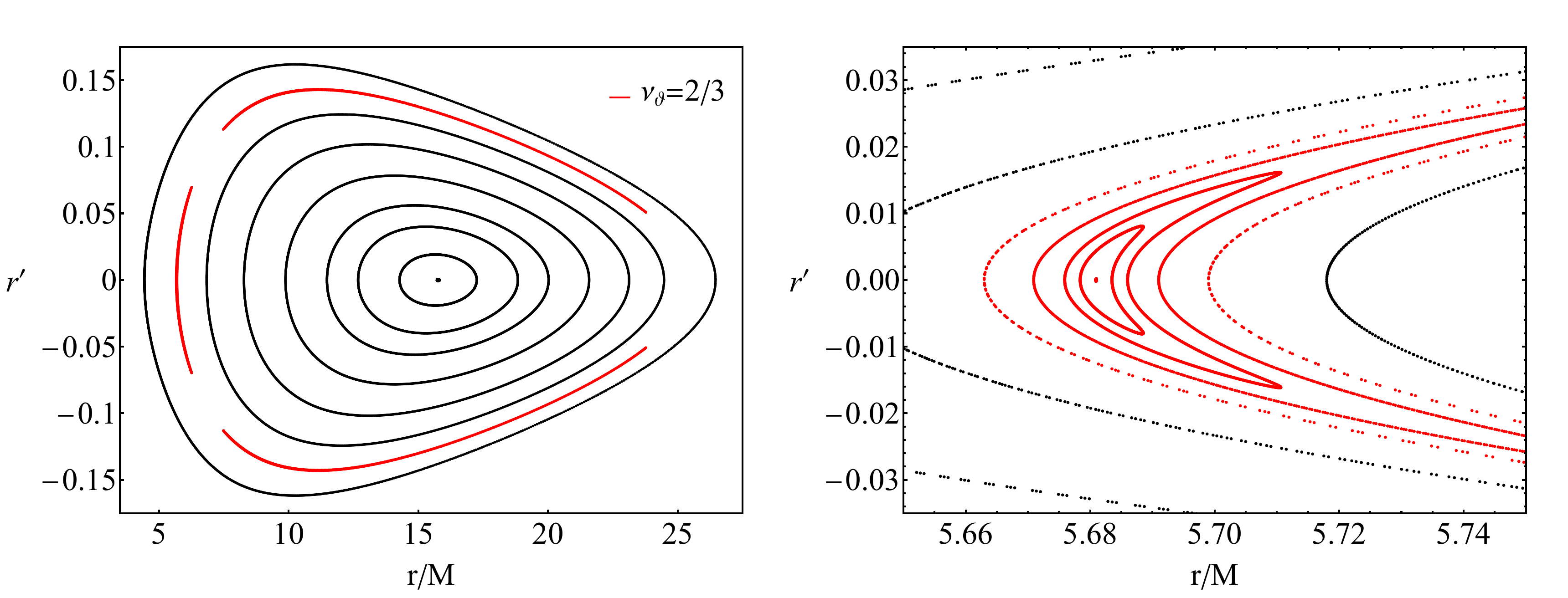}
	\caption{Left: Poincar\'e map of geodesics in HT spacetime with $E/\mu=0.97$, $L_z/\mu=3.5M$, $\chi=0.3$, $\delta q=1$. The black curves correspond to typical KAM curves while the red ones belong to the the resonant KAM curves that surround the period stable point. Right: Zoom in to the leftmost island from Left.}
	\label{fig:Poincare_map}
\end{figure*}

\begin{figure}[t]\centering
	\includegraphics[scale=0.31]{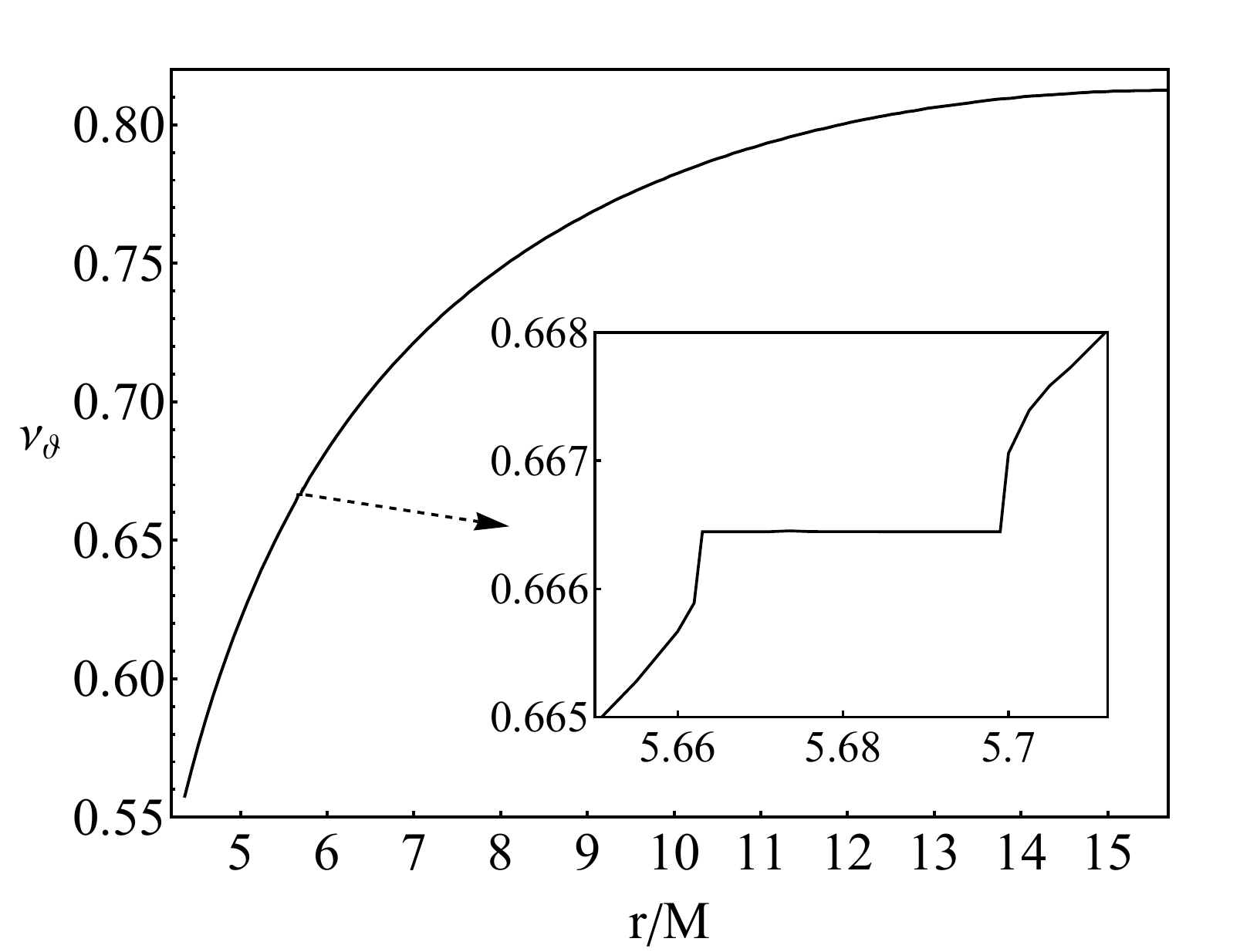}
	\caption{Rotation curve of geodesics in HT metric from Fig. \ref{fig:potential_CZV}. The inset zooms in the region where the $2/3$ plateau is clearly visible.}
	\label{fig:rotation_curve_case}
\end{figure}

The two tools we will utilize in this work to conclude if chaos exists also in particle orbits, besides photon ones \cite{Kostaros:2021usv}, are the Poincar\'e map and rotation curve \cite{Contopoulos_book}. Poincar\'e maps are formally defined as an accumulation of KAM curves for successive geodesics, by fixing all parameters and varying one of them, such as $r$ or $\theta$. The definition explicitly requires that the intersection through the chosen surface of section to be one directional and perpendicular to the surface. By sketching a high resolution Poincar\'e map, one can find resonant islands and Birkhoff chains depending on the system and the deformation from Kerr used. In HT spacetime, which we will operate, the deformation parameter is $\delta q$, but nevertheless we will also explore the dependence of the islands width with respect to all the spacetime parameters and particle geodesics.

The rotation number is helpful to find regions where islands of stability exist. We calculate it by tracking the angle $\vartheta$ between successive intersections of KAM curves, relative to the fixed periodic central point of the Poincar\'e map. The rotation number is then defined as the series of all angles $\vartheta$ measured between one directional consecutive intersections, i.e.~\cite{Contopoulos_book}
\begin{equation}\label{rotation}
	\nu_\vartheta=\frac{1}{2 \pi N}\sum_{i=1}^{N}\vartheta_i,
\end{equation}
where $N$ is the number of angles measured. When $N\rightarrow\infty$, Eq.~\eqref{rotation} asymptotes to $\nu_\vartheta=\omega_r/\omega_\theta$. Calculating consecutive rotation numbers for different initial conditions of orbits, by smoothly varying one of the parameters of the system while keeping the rest fixed, leads to a rotation curve. The more angles measured, which corresponds to more evolution time of the geodesic, the better we approximate the rotation number. Empirical tests have shown that the rotation number's accuracy is inversely proportional to the number of intersections through the surface of section, i.e. evolving an orbit for $10^4$ cycles, which leads to $\sim10^4$ intersections, estimates the rotation number with order $\mathcal{O}(10^{-4})$ accuracy (four decimal points). We stress that this is an empirical law that seems to hold for gravitational systems and is not a systematic rule for all possible scenarios.

Integrable systems have monotonic rotation curves, while nonintegrable ones display discontinuities in the monotonicity through the formation of transient plateaus with non-zero widths when geodesics traverse resonant islands (the rotation number is shared throughout the island). Inflection points can also appear when trajectories pass through unstable periodic points. Nevertheless, by varying the initial conditions, we can drive the orbit through any island and obtain a plateau.

\section{Potential and Curves of Zero Velocity}

\begin{figure*}[t]\centering
	\includegraphics[scale=0.24]{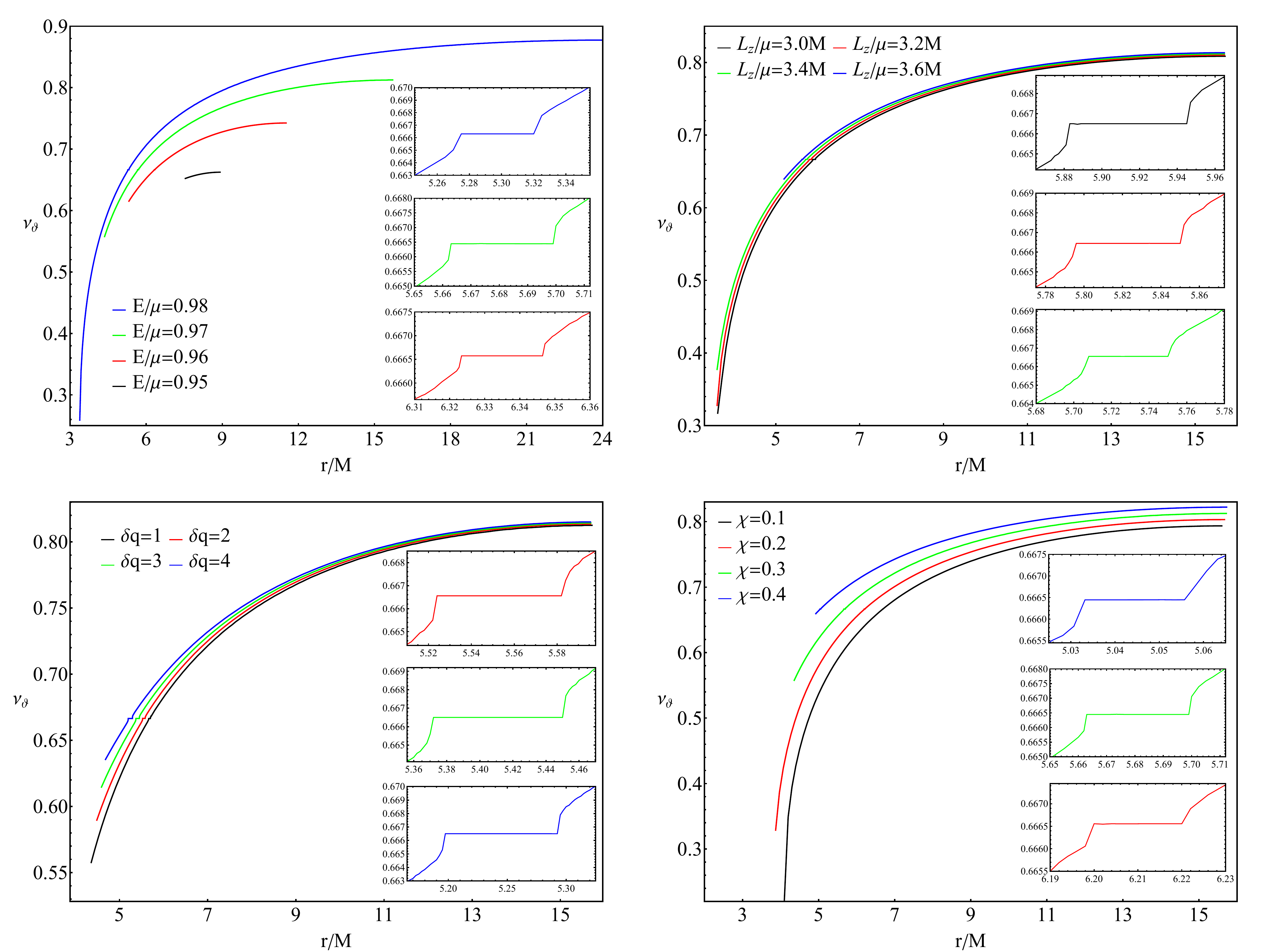}
	\caption{Top left: Rotation curves of geodesics in HT metric with fixed $L_z/\mu=3.5M$, $\delta q=1$, $\chi=0.3$ and varying $E/\mu$. The zoom-in insets portray the plateaus appearing in $2/3$ islands for some of the cases. Top right: Same us top left with fixed $E/\mu=0.97$, $\delta q=1$, $\chi=0.3$ and varying $L_z/\mu$. Bottom left: Same as top left with fixed $E/\mu=0.97$, $L_z/\mu=3.5M$, $\chi=0.3$ and varying $\delta q$. Bottom right: Same as top left with fixed $E/\mu=0.97$, $L_z/\mu=3.5M$, $\delta q=1$ and varying $\chi$.}
	\label{fig:rotation_curves_all}
\end{figure*}

To gain an understanding of the region of bounded motion for test particles in the HT metric, one can begin by comprehending the effective potential $V_\text{eff}$ and the corresponding CZVs for various combinations of parameters. This study should provide information about the regions where bound motion is permitted, as well as the motion's form. For instance, it could clarify whether the CZV is entirely closed and geodesics starting inside it can never escape, or if it is open toward the surface of the compact object and can lead to plunging orbits. Since the HT metric is symmetric in the equatorial plane, it is enough to plot the potential for $\theta=\pi/2$, where it achieves its deepest valley (if it exists).

Fig. \ref{fig:potential_CZV} demonstrates the behavior of both the equatorial potential and the CZVs for varying $E$ and $L_z$. We observe that as the energy of the particle increases, the potential forms a bigger valley and this leads to a larger volume in the CZV where bound motion can occur. Notice that for the particular choice of fixed $L_z$ in Fig. \ref{fig:potential_CZV}, all CZV are closed so there are no plunging orbits that begin from inside the region of bound motion and abide to the appropriate initial conditions and constraint equation. 
On the contrary, we may fix $E$ and vary $L_z$. Then by increasing $L_z$ we observe that the potential changes from one that intersects $V_\text{eff}=0$ only once to one that forms a valley. Therefore, the increment of $L_z$ changes the nature of the corresponding CZVs from open, that allow plunging orbits, to closed where only bound motion occurs. It is curious that, in the case of the largest $L_z$ presented in Figure \ref{fig:potential_CZV}, there is a secondary region very close to the compact object where only plunging orbits occur. Therefore, we will avoid these regimes in phase space when they appear.

\section{Results}

To understand the appearance of indirect chaos on a Poincar\'e map, we present a typical map of the HT metric with a quadrupole deformation of $\delta q=1$ in Figure \ref{fig:Poincare_map}. Although the majority of the curves formed surround the central periodic point in an ordered fashion, as if the system were integrable, a thorough examination reveals the presence of a $2/3$ resonant island (highlighted in red). Upon closer inspection of the leftmost island of the Birkhoff chain, we observe the encapsulating structure of the island and how the resonant KAM curves surround the stable periodic point of the resonance rather than that of the map. This is an indirect manifestation of chaos, and the system being studied is nonintegrable, at least up to second order in the rotation rate, denoted as $\mathcal{O}(\Omega^2)$. In a recent study \cite{Kostaros:2021usv}, the authors analyzed photon orbits for the HT metric at the next leading order, i.e., $\mathcal{O}(\Omega^3)$, and still observed chaotic effects. Therefore, we do not expect that including even higher orders of rotation will make the system integrable, as has been observed in Kerr spacetime \cite{Deich:2022vna}. Figure \ref{fig:rotation_curve_case} completes our basic demonstration of nonintegrability. For the same parameters as in Fig. \ref{fig:Poincare_map}, we sketch the rotation curve from the left edge of the CZV up to the central point of the Poincar\'e map. We find that the existence of the $2/3$ island breaks the monotonicity of the rotation curve and at $\nu_\vartheta=2/3$ it presents a plateau where the rotation number remains constant for a range of initial conditions $r(0)/M$. To determine the width of the island, we run geodesics near the entrance and exit of the plateau and extract the initial conditions that result in a $2/3$ island in the Poincar\'e map. This ensures that we are inside the island upon entry and just before exit.

\begin{figure*}[b!]\centering
	\includegraphics[scale=0.34]{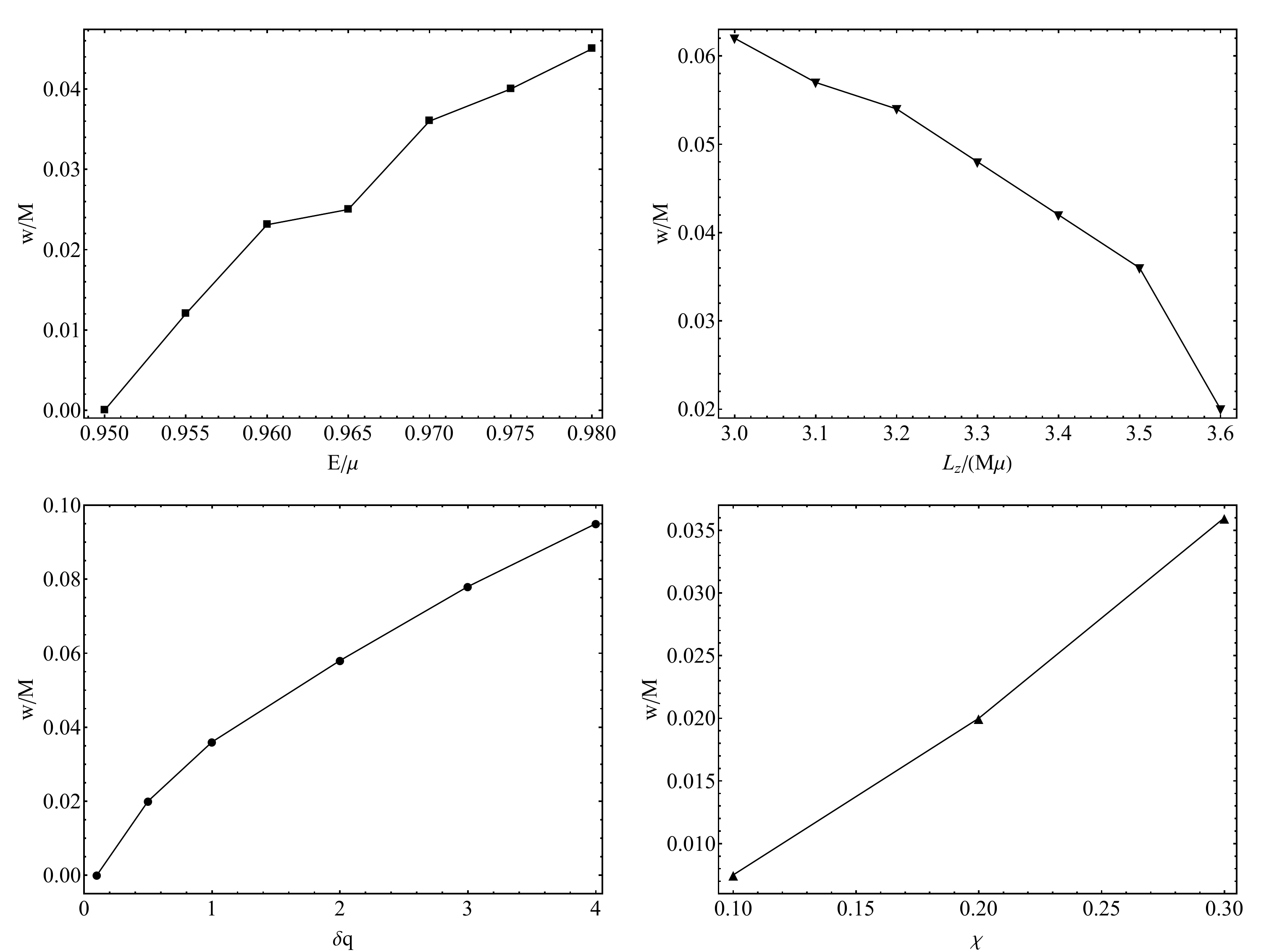}
	\caption{Top left: $2/3$ resonant island widths $w/M$ of geodesics in HT metric with fixed $L_z/\mu=3.5M$, $\delta q=1$, $\chi=0.3$ and varying $E/\mu$. Top right: Same us top left with fixed $E/\mu=0.97$, $\delta q=1$, $\chi=0.3$ and varying $L_z/\mu$. Bottom left: Same as top left with fixed $E/\mu=0.97$, $L_z/\mu=3.5M$, $\chi=0.3$ and varying $\delta q$. Bottom right: Same as top left with fixed $E/\mu=0.97$, $L_z/\mu=3.5M$, $\delta q=1$ and varying $\chi$.}
	\label{fig:width}
\end{figure*}

\subsection{Varying $E/\mu$}

As demonstrated in Fig. \ref{fig:potential_CZV}, the relationship between the potential and CZV is proportional to the magnitude of $E$. The rotation curves displayed in Fig. \ref{fig:rotation_curves_all} (top left) exhibit the anticipated behavior, whereby an increase in the particle's energy results in rotation curves spanning a greater range of $r(0)/M$ due to the expansion of the CZV volume. If $E/\mu=0.95$, the $2/3$ resonance is not satisfied by any bound geodesic. This can be misleading and result in incorrect assumptions about integrability since there are no plateaus. However, changing $E$ will eventually disrupt the smoothness of the rotation curve. The insets in Fig. \ref{fig:rotation_curves_all} (top left) zoom in on the regions where the plateau occurs. We observe that for the cases presented, the width of the plateau's width increases with $E$. In fact, Fig. \ref{fig:width} (top left) shows the dependence of the width of the $2/3$ island on $E$, which clearly demonstrates that the width increases when the particle is initialized with more energy.

\subsection{Varying $L_z/\mu$}

According to Fig. \ref{fig:potential_CZV}, if we increase $L_z$, then the width of the island is expected to decrease as the volume of the  CZV is reduced, even though it becomes closed. This means that the geodesic has access to fewer initial conditions in general. In this case, the dependence of the potential and CZV is inversely proportional to the size of $L_z$. The rotation curves in Fig. \ref{fig:rotation_curves_all} (top right) demonstrate this behavior, as increasing the momentum of the particles leads to rotation curves that cover smaller ranges of $r(0)/M$. The insets shown in the top right of Fig. \ref{fig:rotation_curves_all} zoom onto the regions where the plateau occurs. For the cases presented, the plateau widths decrease as $L_z$ increases. Figure \ref{fig:width} (top right) confirms this trend, as the width of the $2/3$ island decreases with increasing $L_z$. Thus, as the particle's $z$-component of angular momentum is reduced, the island's width increases. However, from Fig. \ref{fig:width}, it is also apparent that by decreasing $L_z$ even further while keeping the other parameters constant, we can reach islands with widths of around $\mathcal{O}(10^{-1}M)$, which are quite large for slowly-rotating compact objects. This is especially true when compared to rapidly-rotating non-Kerr and bumpy spacetimes \cite{Destounis:2020kss,Destounis:2021mqv,Destounis:2021rko}. It should be noted that rotating boson stars, depending on their compactness, can reach unprecedented levels of island widths of order $\mathcal{O}(M)$ \cite{Destounis:2023khj}.

\subsection{Varying $\chi$}

The expectation that the increase in spin will expand the width of the island has been thoroughly demonstrated in \cite{Lukes-Gerakopoulos:2010ipp}. Our results align with this analysis even though we are limited in how large the HT spin can be due to the perturbative nature of the metric. Fig. \ref{fig:rotation_curves_all} (bottom right) shows exactly what we expected: the increase in $\chi$ results in larger $2/3$ islands but smaller CZVs as the range of $r(0)/M$ is reduced. Figure \ref{fig:width} is consistent with this and other analyses found in the literature \cite{Lukes-Gerakopoulos:2010ipp, Lukes-Gerakopoulos:2013gwa,Lukes-Gerakopoulos:2014dpa}.

\subsection{Varying $\delta q$}

The final and most crucial parameter that we have varied is the quadrupole deformation $\delta q$. We kept $\delta q=1$ as a reference and increased it accordingly. Figures \ref{fig:rotation_curves_all} and \ref{fig:width} (bottom left) show that the further we deviate from the Kerr mass quadrupole, the more noticeable the chaotic phenomenon becomes. This finding aligns with non-Kerr spacetimes as reported in \cite{Lukes-Gerakopoulos:2010ipp}. Island widths increase significantly as the deformation increases, reaching widths on the order of $\mathcal{O}(10^{-1}M)$ for a sufficiently large $\delta q$. This is a fascinating result, particularly considering that the spacetime is slowly rotating. However, it should be noted that the HT metric is only an approximate solution to the field equations, and the quadrupole deformation must be very large for prominent chaotic phenomena to occur. Based on the analysis presented above, it is clear that particle geodesics around the HT metric are nonintegrable to at least $\mathcal{O}(\Omega^2)$ spin, and possibly even higher orders in $\Omega$ (as discussed in \cite{Kostaros:2021usv}) but this is just speculation since perturbative solutions have to be treaded lightly.

\section{Conclusions and Discussion}

In this work, we have studied the geodesics of massive particles around the HT metric which can potentially describe slowly-rotating massive stars, white dwarfs and NSs. We have explored the available parameter space of perturbative HT solutions up to $\mathcal{O}(\Omega^2)$ in rotation order by evolving non-spherical, generic particle orbits within the allowed regions of bound motion. Our analysis not only reveals how the structure of the metric and the region of bound geodesics changes when we vary the parameters of spacetime and the particle but also confirms, numerically, the fact that geodesics are nonintegrable. Nonintegrability has also been found for photon orbits in HT spacetime \cite{Kostaros:2021usv} for $\mathcal{O}(\Omega^2)$ and $\mathcal{O}(\Omega^3)$ in spin, where for the perturbative HT metric up to $\mathcal{O}(\Omega^3)$ the chaotic effects are amplified with respect to those in $\mathcal{O}(\Omega^2)$. This means that the evolution of particles is not deterministic, and chaotic effects occur. 

We demonstrated how chaotic features arise using two tools: the Poincar\'e map and the corresponding rotation curve of successive geodesics with varying initial conditions. Poincar\'e maps reveal that most of the phase space seems to generate integrable geodesics around the compact object, with KAM curves enclosing the central periodic point of the map, until a strong resonance is reached. In these regions, KAM curves break down into periodic stable orbits surrounded by resonant KAM curves and encapsulated by islands of stability. The indirect indication of chaos is imprinted and quantified through the rotation number, that is shared with that of the central periodic point (in our case, $\nu_\vartheta=2/3$) for all geodesics that occupy the island. This phenomenon does not appear in integrable spacetimes like Kerr. The shared rotation number translates to a plateau in the rotation curve. We expect other subdominant resonances to exhibit the same features, but only when the deformation parameter is much larger and higher perturbative orders in the rotation are taken into account, which introduce further deformation parameters. Alternatively, the deformation can be chosen to be oblate, meaning $\delta q<0$. However, in \cite{Kostaros:2021usv} it is argued that oblate deformations do not generate chaotic effects as strong as those for prolate deformations. We did not explore extremely large mass-quadrupole deformations to avoid deviating too much from General Relativity and to satisfy the pathological bounds of the HT metric \cite{DeFalco:2021nfk}.

\noindent Our extensive, and quantitative, analysis reveals the following: 

(i) As we increase the particle's energy ($E$), the volume of CZVs increases, allowing more geodesics to be bound around the HT spacetime. Consequently, the widths of $2/3$ resonant islands increase with $E$.

(ii) Increasing the particle's $z$-component of angular momentum has the opposite effect to that in (i). CZVs change from an open configuration, which allows for plunging orbits, to a closed and smaller-volume one. Thus, the island widths decrease with $L_z$.

(iii) The widths of $2/3$ resonant islands are proportional to both spacetime parameters, namely quadrupole deformation ($\delta q$) and spin ($\chi$). Specifically, for a suitable combination of large $E$, $\delta q$, and spin, as well as small $L_z$, island widths of order $\mathcal{O}(w/M)\gtrsim 10^{-1}$, as in \cite{Lukes-Gerakopoulos:2010ipp}, can be achieved. This is of interest since the HT metric is slowly-rotating and perturbative to second order in spin.

Our results confirm the nonintegrability of the HT metric, even at $\mathcal{O}(\Omega^2)$ in rotation rate, and quantify how all parameters involved in test-particle generic geodesics affect the orbital motion and chaotic regions. A potential extension of our work would be to perform a similar analysis but for higher order perturbative HT metrics, by using the perturbative solutions in $\mathcal{O}(\Omega^3)$ and $\mathcal{O}(\Omega^4)$ \cite{Glampedakis:2017cgd,Yagi:2014bxa}, in order to further quantify the extent to which chaotic regions and resonant islands are enhanced, due to the further deformations that will be introduced to the multipolar structure of the HT metric. Nevertheless, we expect nonintegrability and chaos to be a generic feature of HT spacetimes. To exit the realm of geodesics around perturbative solutions that are being used as if they are those of an exact spacetime, a future direction could be the use of numerical methods, as outlined in \cite{Stergioulas:1994ea,Stergioulas:2003yp,Berti:2003nb}, in order to evolve particle orbits around numerically generated solutions, such as rapidly rotating NSs, as done in \cite{Destounis:2023khj} for boson stars, which is a much more demanding task and falls out of the scope of the current analysis. Finally, it may be worth exploring whether resonant islands and chaotic effects can potentially be connected to QPOs of accreting discs around NSs and non-Kerr BHs. The ultimate goal would be to estimate the mass, angular momentum, and multipole moments of accreting compact objects associated with low-mass X-ray binaries in order to distinguish between BHs and NSs \cite{Boshkayev:2014mua,Boshkayev:2015ARep...59..441B,Boshkayev:2016epd}.


\section*{Declarations}
We the authors hereby declare that there are no competing interests of financial or personal nature. Ethical approval for this work is not applicable. All authors have contributed equally for the execution of calculations and the presentation of results, as well as the creation of this manuscript. The data that support the findings of this study are available from the corresponding author upon reasonable request. 
This work was supported by the DAAD program for the ``promotion of the exchange and scientific cooperation between Greece and Germany IKYDAAD 2022" (57628320). K.D. and K.K. are grateful for hospitality provided by the Department of Physics of the University of Athens, Greece. K.D. further acknowledges financial support provided under the European Union's H2020 ERC, Starting Grant agreement no.~DarkGRA--757480 and the MIUR PRIN and FARE programmes (GW-NEXT, CUP: B84I20000100001). The authors would like to, finally, warmly thank Prof. G. Pappas, Prof. K. Glampedakis, Prof. Th. Apostolatos, Prof. P. Pani, Dr. A. Eleni and Dr. P. Manoharan for critical comments and helpful discussions.

\bibliography{HT_chaos}

\begin{thebibliography}{113}%
\makeatletter
\providecommand \@ifxundefined [1]{%
 \@ifx{#1\undefined}
}%
\providecommand \@ifnum [1]{%
 \ifnum #1\expandafter \@firstoftwo
 \else \expandafter \@secondoftwo
 \fi
}%
\providecommand \@ifx [1]{%
 \ifx #1\expandafter \@firstoftwo
 \else \expandafter \@secondoftwo
 \fi
}%
\providecommand \natexlab [1]{#1}%
\providecommand \enquote  [1]{``#1''}%
\providecommand \bibnamefont  [1]{#1}%
\providecommand \bibfnamefont [1]{#1}%
\providecommand \citenamefont [1]{#1}%
\providecommand \href@noop [0]{\@secondoftwo}%
\providecommand \href [0]{\begingroup \@sanitize@url \@href}%
\providecommand \@href[1]{\@@startlink{#1}\@@href}%
\providecommand \@@href[1]{\endgroup#1\@@endlink}%
\providecommand \@sanitize@url [0]{\catcode `\\12\catcode `\$12\catcode
  `\&12\catcode `\#12\catcode `\^12\catcode `\_12\catcode `\%12\relax}%
\providecommand \@@startlink[1]{}%
\providecommand \@@endlink[0]{}%
\providecommand \url  [0]{\begingroup\@sanitize@url \@url }%
\providecommand \@url [1]{\endgroup\@href {#1}{\urlprefix }}%
\providecommand \urlprefix  [0]{URL }%
\providecommand \Eprint [0]{\href }%
\providecommand \doibase [0]{https://doi.org/}%
\providecommand \selectlanguage [0]{\@gobble}%
\providecommand \bibinfo  [0]{\@secondoftwo}%
\providecommand \bibfield  [0]{\@secondoftwo}%
\providecommand \translation [1]{[#1]}%
\providecommand \BibitemOpen [0]{}%
\providecommand \bibitemStop [0]{}%
\providecommand \bibitemNoStop [0]{.\EOS\space}%
\providecommand \EOS [0]{\spacefactor3000\relax}%
\providecommand \BibitemShut  [1]{\csname bibitem#1\endcsname}%
\let\auto@bib@innerbib\@empty
\bibitem [{\citenamefont {Taylor}\ \emph {et~al.}(1979)\citenamefont {Taylor},
  \citenamefont {Fowler},\ and\ \citenamefont {McCulloch}}]{Taylor:1979zz}%
  \BibitemOpen
  \bibfield  {author} {\bibinfo {author} {\bibfnamefont {J.~H.}\ \bibnamefont
  {Taylor}}, \bibinfo {author} {\bibfnamefont {L.~A.}\ \bibnamefont {Fowler}},\
  and\ \bibinfo {author} {\bibfnamefont {P.~M.}\ \bibnamefont {McCulloch}},\
  }\bibfield  {title} {\bibinfo {title} {{Measurements of general relativistic
  effects in the binary pulsar PSR 1913+16}},\ }\href
  {https://doi.org/10.1038/277437a0} {\bibfield  {journal} {\bibinfo  {journal}
  {Nature}\ }\textbf {\bibinfo {volume} {277}},\ \bibinfo {pages} {437}
  (\bibinfo {year} {1979})}\BibitemShut {NoStop}%
\bibitem [{\citenamefont {van~der Klis}(2000)}]{vanderKlis:2000ca}%
  \BibitemOpen
  \bibfield  {author} {\bibinfo {author} {\bibfnamefont {M.}~\bibnamefont
  {van~der Klis}},\ }\bibfield  {title} {\bibinfo {title} {{Millisecond
  oscillations in x-ray binaries}},\ }\href
  {https://doi.org/10.1146/annurev.astro.38.1.717} {\bibfield  {journal}
  {\bibinfo  {journal} {Ann. Rev. Astron. Astrophys.}\ }\textbf {\bibinfo
  {volume} {38}},\ \bibinfo {pages} {717} (\bibinfo {year} {2000})},\ \Eprint
  {https://arxiv.org/abs/astro-ph/0001167} {arXiv:astro-ph/0001167}
  \BibitemShut {NoStop}%
\bibitem [{\citenamefont {Kluzniak}\ and\ \citenamefont
  {Abramowicz}(2003)}]{Kluzniak:2003ei}%
  \BibitemOpen
  \bibfield  {author} {\bibinfo {author} {\bibfnamefont {W.}~\bibnamefont
  {Kluzniak}}\ and\ \bibinfo {author} {\bibfnamefont {M.}~\bibnamefont
  {Abramowicz}},\ }\bibfield  {title} {\bibinfo {title} {{General Relativity
  and Gravitation. Millisecond oscillators in accreting neutron stars and black
  holes}},\ }in\ \href@noop {} {\emph {\bibinfo {booktitle} {{12th Workshop on
  General Relativity and Gravitation}}}}\ (\bibinfo {year} {2003})\ \Eprint
  {https://arxiv.org/abs/astro-ph/0304345} {arXiv:astro-ph/0304345}
  \BibitemShut {NoStop}%
\bibitem [{\citenamefont {{Lewin}}\ and\ \citenamefont {{van der
  Klis}}(2006)}]{vanderKlis:2006}%
  \BibitemOpen
  \bibfield  {author} {\bibinfo {author} {\bibfnamefont {W.~H.~G.}\
  \bibnamefont {{Lewin}}}\ and\ \bibinfo {author} {\bibfnamefont
  {M.}~\bibnamefont {{van der Klis}}},\ }\href@noop {} {\emph {\bibinfo {title}
  {{Compact Stellar X-ray Sources}}}},\ Vol.~\bibinfo {volume} {39}\ (\bibinfo
  {year} {2006})\BibitemShut {NoStop}%
\bibitem [{\citenamefont {Poynting}(1903)}]{Poynting:10.1093/mnras/64.1.1a}%
  \BibitemOpen
  \bibfield  {author} {\bibinfo {author} {\bibfnamefont {J.~H.}\ \bibnamefont
  {Poynting}},\ }\bibfield  {title} {\bibinfo {title} {{“Radiation in the
  Solar System; its Effect on Temperature and its Pressure on Small
  Bodies.”}},\ }\href {https://doi.org/10.1093/mnras/64.1.1a} {\bibfield
  {journal} {\bibinfo  {journal} {Monthly Notices of the Royal Astronomical
  Society}\ }\textbf {\bibinfo {volume} {64}},\ \bibinfo {pages} {1a} (\bibinfo
  {year} {1903})},\ \Eprint
  {https://arxiv.org/abs/https://academic.oup.com/mnras/article-pdf/64/1/1a/3438550/mnras64-0001a.pdf}
  {https://academic.oup.com/mnras/article-pdf/64/1/1a/3438550/mnras64-0001a.pdf}
  \BibitemShut {NoStop}%
\bibitem [{\citenamefont {Robertson}\ and\ \citenamefont
  {Russell}(1937)}]{Robertson:10.1093/mnras/97.6.423}%
  \BibitemOpen
  \bibfield  {author} {\bibinfo {author} {\bibfnamefont {H.~P.}\ \bibnamefont
  {Robertson}}\ and\ \bibinfo {author} {\bibfnamefont {H.~N.}\ \bibnamefont
  {Russell}},\ }\bibfield  {title} {\bibinfo {title} {{Dynamical Effects of
  Radiation in the Solar System}},\ }\href
  {https://doi.org/10.1093/mnras/97.6.423} {\bibfield  {journal} {\bibinfo
  {journal} {Monthly Notices of the Royal Astronomical Society}\ }\textbf
  {\bibinfo {volume} {97}},\ \bibinfo {pages} {423} (\bibinfo {year} {1937})},\
  \Eprint
  {https://arxiv.org/abs/https://academic.oup.com/mnras/article-pdf/97/6/423/3169085/mnras97-0423.pdf}
  {https://academic.oup.com/mnras/article-pdf/97/6/423/3169085/mnras97-0423.pdf}
  \BibitemShut {NoStop}%
\bibitem [{\citenamefont {Akiyama}\ \emph {et~al.}(2019)\citenamefont {Akiyama}
  \emph {et~al.}}]{EventHorizonTelescope:2019dse}%
  \BibitemOpen
  \bibfield  {author} {\bibinfo {author} {\bibfnamefont {K.}~\bibnamefont
  {Akiyama}} \emph {et~al.} (\bibinfo {collaboration} {Event Horizon
  Telescope}),\ }\bibfield  {title} {\bibinfo {title} {{First M87 Event Horizon
  Telescope Results. I. The Shadow of the Supermassive Black Hole}},\ }\href
  {https://doi.org/10.3847/2041-8213/ab0ec7} {\bibfield  {journal} {\bibinfo
  {journal} {Astrophys. J. Lett.}\ }\textbf {\bibinfo {volume} {875}},\
  \bibinfo {pages} {L1} (\bibinfo {year} {2019})},\ \Eprint
  {https://arxiv.org/abs/1906.11238} {arXiv:1906.11238 [astro-ph.GA]}
  \BibitemShut {NoStop}%
\bibitem [{\citenamefont {Akiyama}\ \emph {et~al.}(2022)\citenamefont {Akiyama}
  \emph {et~al.}}]{EventHorizonTelescope:2022wkp}%
  \BibitemOpen
  \bibfield  {author} {\bibinfo {author} {\bibfnamefont {K.}~\bibnamefont
  {Akiyama}} \emph {et~al.} (\bibinfo {collaboration} {Event Horizon
  Telescope}),\ }\bibfield  {title} {\bibinfo {title} {{First Sagittarius A*
  Event Horizon Telescope Results. I. The Shadow of the Supermassive Black Hole
  in the Center of the Milky Way}},\ }\href
  {https://doi.org/10.3847/2041-8213/ac6674} {\bibfield  {journal} {\bibinfo
  {journal} {Astrophys. J. Lett.}\ }\textbf {\bibinfo {volume} {930}},\
  \bibinfo {pages} {L12} (\bibinfo {year} {2022})}\BibitemShut {NoStop}%
\bibitem [{\citenamefont {Abbott}\ \emph {et~al.}(2016)\citenamefont {Abbott}
  \emph {et~al.}}]{LIGOScientific:2016aoc}%
  \BibitemOpen
  \bibfield  {author} {\bibinfo {author} {\bibfnamefont {B.~P.}\ \bibnamefont
  {Abbott}} \emph {et~al.} (\bibinfo {collaboration} {LIGO Scientific,
  Virgo}),\ }\bibfield  {title} {\bibinfo {title} {{Observation of
  Gravitational Waves from a Binary Black Hole Merger}},\ }\href
  {https://doi.org/10.1103/PhysRevLett.116.061102} {\bibfield  {journal}
  {\bibinfo  {journal} {Phys. Rev. Lett.}\ }\textbf {\bibinfo {volume} {116}},\
  \bibinfo {pages} {061102} (\bibinfo {year} {2016})},\ \Eprint
  {https://arxiv.org/abs/1602.03837} {arXiv:1602.03837 [gr-qc]} \BibitemShut
  {NoStop}%
\bibitem [{\citenamefont {Abbott}\ \emph
  {et~al.}(2021{\natexlab{a}})\citenamefont {Abbott} \emph
  {et~al.}}]{LIGOScientific:2021djp}%
  \BibitemOpen
  \bibfield  {author} {\bibinfo {author} {\bibfnamefont {R.}~\bibnamefont
  {Abbott}} \emph {et~al.} (\bibinfo {collaboration} {LIGO Scientific, VIRGO,
  KAGRA}),\ }\bibfield  {title} {\bibinfo {title} {{GWTC-3: Compact Binary
  Coalescences Observed by LIGO and Virgo During the Second Part of the Third
  Observing Run}},\ }\href@noop {} {\  (\bibinfo {year}
  {2021}{\natexlab{a}})},\ \Eprint {https://arxiv.org/abs/2111.03606}
  {arXiv:2111.03606 [gr-qc]} \BibitemShut {NoStop}%
\bibitem [{\citenamefont {Abbott}\ \emph
  {et~al.}(2017{\natexlab{a}})\citenamefont {Abbott} \emph
  {et~al.}}]{LIGOScientific:2017vwq}%
  \BibitemOpen
  \bibfield  {author} {\bibinfo {author} {\bibfnamefont {B.~P.}\ \bibnamefont
  {Abbott}} \emph {et~al.} (\bibinfo {collaboration} {LIGO Scientific,
  Virgo}),\ }\bibfield  {title} {\bibinfo {title} {{GW170817: Observation of
  Gravitational Waves from a Binary Neutron Star Inspiral}},\ }\href
  {https://doi.org/10.1103/PhysRevLett.119.161101} {\bibfield  {journal}
  {\bibinfo  {journal} {Phys. Rev. Lett.}\ }\textbf {\bibinfo {volume} {119}},\
  \bibinfo {pages} {161101} (\bibinfo {year} {2017}{\natexlab{a}})},\ \Eprint
  {https://arxiv.org/abs/1710.05832} {arXiv:1710.05832 [gr-qc]} \BibitemShut
  {NoStop}%
\bibitem [{\citenamefont {Abbott}\ \emph
  {et~al.}(2021{\natexlab{b}})\citenamefont {Abbott} \emph
  {et~al.}}]{LIGOScientific:2021qlt}%
  \BibitemOpen
  \bibfield  {author} {\bibinfo {author} {\bibfnamefont {R.}~\bibnamefont
  {Abbott}} \emph {et~al.} (\bibinfo {collaboration} {LIGO Scientific, KAGRA,
  VIRGO}),\ }\bibfield  {title} {\bibinfo {title} {{Observation of
  Gravitational Waves from Two Neutron Star\textendash{}Black Hole
  Coalescences}},\ }\href {https://doi.org/10.3847/2041-8213/ac082e} {\bibfield
   {journal} {\bibinfo  {journal} {Astrophys. J. Lett.}\ }\textbf {\bibinfo
  {volume} {915}},\ \bibinfo {pages} {L5} (\bibinfo {year}
  {2021}{\natexlab{b}})},\ \Eprint {https://arxiv.org/abs/2106.15163}
  {arXiv:2106.15163 [astro-ph.HE]} \BibitemShut {NoStop}%
\bibitem [{\citenamefont {Barack}\ \emph {et~al.}(2019)\citenamefont {Barack}
  \emph {et~al.}}]{Barack:2018yly}%
  \BibitemOpen
  \bibfield  {author} {\bibinfo {author} {\bibfnamefont {L.}~\bibnamefont
  {Barack}} \emph {et~al.},\ }\bibfield  {title} {\bibinfo {title} {{Black
  holes, gravitational waves and fundamental physics: a roadmap}},\ }\href
  {https://doi.org/10.1088/1361-6382/ab0587} {\bibfield  {journal} {\bibinfo
  {journal} {Class. Quant. Grav.}\ }\textbf {\bibinfo {volume} {36}},\ \bibinfo
  {pages} {143001} (\bibinfo {year} {2019})},\ \Eprint
  {https://arxiv.org/abs/1806.05195} {arXiv:1806.05195 [gr-qc]} \BibitemShut
  {NoStop}%
\bibitem [{\citenamefont {Barausse}\ \emph {et~al.}(2014)\citenamefont
  {Barausse}, \citenamefont {Cardoso},\ and\ \citenamefont
  {Pani}}]{Barausse:2014tra}%
  \BibitemOpen
  \bibfield  {author} {\bibinfo {author} {\bibfnamefont {E.}~\bibnamefont
  {Barausse}}, \bibinfo {author} {\bibfnamefont {V.}~\bibnamefont {Cardoso}},\
  and\ \bibinfo {author} {\bibfnamefont {P.}~\bibnamefont {Pani}},\ }\bibfield
  {title} {\bibinfo {title} {{Can environmental effects spoil precision
  gravitational-wave astrophysics?}},\ }\href
  {https://doi.org/10.1103/PhysRevD.89.104059} {\bibfield  {journal} {\bibinfo
  {journal} {Phys. Rev. D}\ }\textbf {\bibinfo {volume} {89}},\ \bibinfo
  {pages} {104059} (\bibinfo {year} {2014})},\ \Eprint
  {https://arxiv.org/abs/1404.7149} {arXiv:1404.7149 [gr-qc]} \BibitemShut
  {NoStop}%
\bibitem [{\citenamefont {Cardoso}\ \emph
  {et~al.}(2022{\natexlab{a}})\citenamefont {Cardoso}, \citenamefont
  {Destounis}, \citenamefont {Duque}, \citenamefont {Macedo},\ and\
  \citenamefont {Maselli}}]{Cardoso:2021wlq}%
  \BibitemOpen
  \bibfield  {author} {\bibinfo {author} {\bibfnamefont {V.}~\bibnamefont
  {Cardoso}}, \bibinfo {author} {\bibfnamefont {K.}~\bibnamefont {Destounis}},
  \bibinfo {author} {\bibfnamefont {F.}~\bibnamefont {Duque}}, \bibinfo
  {author} {\bibfnamefont {R.~P.}\ \bibnamefont {Macedo}},\ and\ \bibinfo
  {author} {\bibfnamefont {A.}~\bibnamefont {Maselli}},\ }\bibfield  {title}
  {\bibinfo {title} {{Black holes in galaxies: Environmental impact on
  gravitational-wave generation and propagation}},\ }\href
  {https://doi.org/10.1103/PhysRevD.105.L061501} {\bibfield  {journal}
  {\bibinfo  {journal} {Phys. Rev. D}\ }\textbf {\bibinfo {volume} {105}},\
  \bibinfo {pages} {L061501} (\bibinfo {year} {2022}{\natexlab{a}})},\ \Eprint
  {https://arxiv.org/abs/2109.00005} {arXiv:2109.00005 [gr-qc]} \BibitemShut
  {NoStop}%
\bibitem [{\citenamefont {Cardoso}\ \emph
  {et~al.}(2022{\natexlab{b}})\citenamefont {Cardoso}, \citenamefont
  {Destounis}, \citenamefont {Duque}, \citenamefont {Panosso~Macedo},\ and\
  \citenamefont {Maselli}}]{Cardoso:2022whc}%
  \BibitemOpen
  \bibfield  {author} {\bibinfo {author} {\bibfnamefont {V.}~\bibnamefont
  {Cardoso}}, \bibinfo {author} {\bibfnamefont {K.}~\bibnamefont {Destounis}},
  \bibinfo {author} {\bibfnamefont {F.}~\bibnamefont {Duque}}, \bibinfo
  {author} {\bibfnamefont {R.}~\bibnamefont {Panosso~Macedo}},\ and\ \bibinfo
  {author} {\bibfnamefont {A.}~\bibnamefont {Maselli}},\ }\bibfield  {title}
  {\bibinfo {title} {{Gravitational Waves from Extreme-Mass-Ratio Systems in
  Astrophysical Environments}},\ }\href
  {https://doi.org/10.1103/PhysRevLett.129.241103} {\bibfield  {journal}
  {\bibinfo  {journal} {Phys. Rev. Lett.}\ }\textbf {\bibinfo {volume} {129}},\
  \bibinfo {pages} {241103} (\bibinfo {year} {2022}{\natexlab{b}})},\ \Eprint
  {https://arxiv.org/abs/2210.01133} {arXiv:2210.01133 [gr-qc]} \BibitemShut
  {NoStop}%
\bibitem [{\citenamefont {Destounis}\ \emph
  {et~al.}(2023{\natexlab{a}})\citenamefont {Destounis}, \citenamefont
  {Kulathingal}, \citenamefont {Kokkotas},\ and\ \citenamefont
  {Papadopoulos}}]{Destounis:2022obl}%
  \BibitemOpen
  \bibfield  {author} {\bibinfo {author} {\bibfnamefont {K.}~\bibnamefont
  {Destounis}}, \bibinfo {author} {\bibfnamefont {A.}~\bibnamefont
  {Kulathingal}}, \bibinfo {author} {\bibfnamefont {K.~D.}\ \bibnamefont
  {Kokkotas}},\ and\ \bibinfo {author} {\bibfnamefont {G.~O.}\ \bibnamefont
  {Papadopoulos}},\ }\bibfield  {title} {\bibinfo {title} {{Gravitational-wave
  imprints of compact and galactic-scale environments in extreme-mass-ratio
  binaries}},\ }\href {https://doi.org/10.1103/PhysRevD.107.084027} {\bibfield
  {journal} {\bibinfo  {journal} {Phys. Rev. D}\ }\textbf {\bibinfo {volume}
  {107}},\ \bibinfo {pages} {084027} (\bibinfo {year} {2023}{\natexlab{a}})},\
  \Eprint {https://arxiv.org/abs/2210.09357} {arXiv:2210.09357 [gr-qc]}
  \BibitemShut {NoStop}%
\bibitem [{\citenamefont {Cheung}\ \emph {et~al.}(2022)\citenamefont {Cheung},
  \citenamefont {Destounis}, \citenamefont {Macedo}, \citenamefont {Berti},\
  and\ \citenamefont {Cardoso}}]{Cheung:2021bol}%
  \BibitemOpen
  \bibfield  {author} {\bibinfo {author} {\bibfnamefont {M.~H.-Y.}\
  \bibnamefont {Cheung}}, \bibinfo {author} {\bibfnamefont {K.}~\bibnamefont
  {Destounis}}, \bibinfo {author} {\bibfnamefont {R.~P.}\ \bibnamefont
  {Macedo}}, \bibinfo {author} {\bibfnamefont {E.}~\bibnamefont {Berti}},\ and\
  \bibinfo {author} {\bibfnamefont {V.}~\bibnamefont {Cardoso}},\ }\bibfield
  {title} {\bibinfo {title} {{Destabilizing the Fundamental Mode of Black
  Holes: The Elephant and the Flea}},\ }\href
  {https://doi.org/10.1103/PhysRevLett.128.111103} {\bibfield  {journal}
  {\bibinfo  {journal} {Phys. Rev. Lett.}\ }\textbf {\bibinfo {volume} {128}},\
  \bibinfo {pages} {111103} (\bibinfo {year} {2022})},\ \Eprint
  {https://arxiv.org/abs/2111.05415} {arXiv:2111.05415 [gr-qc]} \BibitemShut
  {NoStop}%
\bibitem [{\citenamefont {Abbott}\ \emph
  {et~al.}(2017{\natexlab{b}})\citenamefont {Abbott} \emph
  {et~al.}}]{LIGOScientific:2017zic}%
  \BibitemOpen
  \bibfield  {author} {\bibinfo {author} {\bibfnamefont {B.~P.}\ \bibnamefont
  {Abbott}} \emph {et~al.} (\bibinfo {collaboration} {LIGO Scientific, Virgo,
  Fermi-GBM, INTEGRAL}),\ }\bibfield  {title} {\bibinfo {title} {{Gravitational
  Waves and Gamma-rays from a Binary Neutron Star Merger: GW170817 and GRB
  170817A}},\ }\href {https://doi.org/10.3847/2041-8213/aa920c} {\bibfield
  {journal} {\bibinfo  {journal} {Astrophys. J. Lett.}\ }\textbf {\bibinfo
  {volume} {848}},\ \bibinfo {pages} {L13} (\bibinfo {year}
  {2017}{\natexlab{b}})},\ \Eprint {https://arxiv.org/abs/1710.05834}
  {arXiv:1710.05834 [astro-ph.HE]} \BibitemShut {NoStop}%
\bibitem [{\citenamefont {Kerr}(1963)}]{Kerr:1963ud}%
  \BibitemOpen
  \bibfield  {author} {\bibinfo {author} {\bibfnamefont {R.~P.}\ \bibnamefont
  {Kerr}},\ }\bibfield  {title} {\bibinfo {title} {{Gravitational field of a
  spinning mass as an example of algebraically special metrics}},\ }\href
  {https://doi.org/10.1103/PhysRevLett.11.237} {\bibfield  {journal} {\bibinfo
  {journal} {Phys. Rev. Lett.}\ }\textbf {\bibinfo {volume} {11}},\ \bibinfo
  {pages} {237} (\bibinfo {year} {1963})}\BibitemShut {NoStop}%
\bibitem [{\citenamefont {Berti}\ and\ \citenamefont
  {Stergioulas}(2004)}]{Berti:2003nb}%
  \BibitemOpen
  \bibfield  {author} {\bibinfo {author} {\bibfnamefont {E.}~\bibnamefont
  {Berti}}\ and\ \bibinfo {author} {\bibfnamefont {N.}~\bibnamefont
  {Stergioulas}},\ }\bibfield  {title} {\bibinfo {title} {{Approximate matching
  of analytic and numerical solutions for rapidly rotating neutron stars}},\
  }\href {https://doi.org/10.1111/j.1365-2966.2004.07740.x} {\bibfield
  {journal} {\bibinfo  {journal} {Mon. Not. Roy. Astron. Soc.}\ }\textbf
  {\bibinfo {volume} {350}},\ \bibinfo {pages} {1416} (\bibinfo {year}
  {2004})},\ \Eprint {https://arxiv.org/abs/gr-qc/0310061}
  {arXiv:gr-qc/0310061} \BibitemShut {NoStop}%
\bibitem [{\citenamefont {Pappas}(2012)}]{Pappas:2012nt}%
  \BibitemOpen
  \bibfield  {author} {\bibinfo {author} {\bibfnamefont {G.}~\bibnamefont
  {Pappas}},\ }\bibfield  {title} {\bibinfo {title} {{What can quasi-periodic
  oscillations tell us about the structure of the corresponding compact
  objects?}},\ }\href {https://doi.org/10.1111/j.1365-2966.2012.20817.x}
  {\bibfield  {journal} {\bibinfo  {journal} {Mon. Not. Roy. Astron. Soc.}\
  }\textbf {\bibinfo {volume} {422}},\ \bibinfo {pages} {2581} (\bibinfo {year}
  {2012})},\ \Eprint {https://arxiv.org/abs/1201.6071} {arXiv:1201.6071
  [astro-ph.HE]} \BibitemShut {NoStop}%
\bibitem [{\citenamefont {{Manko}}\ and\ \citenamefont
  {{Sibgatullin}}(1993)}]{Manko:1993CQGra..10.1383M}%
  \BibitemOpen
  \bibfield  {author} {\bibinfo {author} {\bibfnamefont {V.~S.}\ \bibnamefont
  {{Manko}}}\ and\ \bibinfo {author} {\bibfnamefont {N.~R.}\ \bibnamefont
  {{Sibgatullin}}},\ }\bibfield  {title} {\bibinfo {title} {{Construction of
  exact solutions of the Einstein-Maxwell equations corresponding to a given
  behaviour of the Ernst potentials on the symmetry axis}},\ }\href
  {https://doi.org/10.1088/0264-9381/10/7/014} {\bibfield  {journal} {\bibinfo
  {journal} {Classical and Quantum Gravity}\ }\textbf {\bibinfo {volume}
  {10}},\ \bibinfo {pages} {1383} (\bibinfo {year} {1993})}\BibitemShut
  {NoStop}%
\bibitem [{\citenamefont {{Manko}}\ \emph {et~al.}(1995)\citenamefont
  {{Manko}}, \citenamefont {{Mart{\'\i}n}},\ and\ \citenamefont
  {{Ruiz}}}]{Manko:1995JMP....36.3063M}%
  \BibitemOpen
  \bibfield  {author} {\bibinfo {author} {\bibfnamefont {V.~S.}\ \bibnamefont
  {{Manko}}}, \bibinfo {author} {\bibfnamefont {J.}~\bibnamefont
  {{Mart{\'\i}n}}},\ and\ \bibinfo {author} {\bibfnamefont {E.}~\bibnamefont
  {{Ruiz}}},\ }\bibfield  {title} {\bibinfo {title} {{Six-parameter solution of
  the Einstein{\textendash}Maxwell equations possessing equatorial symmetry}},\
  }\href {https://doi.org/10.1063/1.531012} {\bibfield  {journal} {\bibinfo
  {journal} {Journal of Mathematical Physics}\ }\textbf {\bibinfo {volume}
  {36}},\ \bibinfo {pages} {3063} (\bibinfo {year} {1995})}\BibitemShut
  {NoStop}%
\bibitem [{\citenamefont {Manko}\ \emph
  {et~al.}(2000{\natexlab{a}})\citenamefont {Manko}, \citenamefont
  {Sanabria-Gomez},\ and\ \citenamefont {Manko}}]{Manko:2000sg}%
  \BibitemOpen
  \bibfield  {author} {\bibinfo {author} {\bibfnamefont {V.~S.}\ \bibnamefont
  {Manko}}, \bibinfo {author} {\bibfnamefont {J.~D.}\ \bibnamefont
  {Sanabria-Gomez}},\ and\ \bibinfo {author} {\bibfnamefont {O.~V.}\
  \bibnamefont {Manko}},\ }\bibfield  {title} {\bibinfo {title} {{Nine
  parameter electrovac metric involving rational functions}},\ }\href
  {https://doi.org/10.1103/PhysRevD.62.044048} {\bibfield  {journal} {\bibinfo
  {journal} {Phys. Rev. D}\ }\textbf {\bibinfo {volume} {62}},\ \bibinfo
  {pages} {044048} (\bibinfo {year} {2000}{\natexlab{a}})}\BibitemShut
  {NoStop}%
\bibitem [{\citenamefont {Manko}\ \emph
  {et~al.}(2000{\natexlab{b}})\citenamefont {Manko}, \citenamefont {Mielke},\
  and\ \citenamefont {Sanabria-Gomez}}]{Manko:2000ud}%
  \BibitemOpen
  \bibfield  {author} {\bibinfo {author} {\bibfnamefont {V.~S.}\ \bibnamefont
  {Manko}}, \bibinfo {author} {\bibfnamefont {E.~W.}\ \bibnamefont {Mielke}},\
  and\ \bibinfo {author} {\bibfnamefont {J.~D.}\ \bibnamefont
  {Sanabria-Gomez}},\ }\bibfield  {title} {\bibinfo {title} {{Exact solution
  for the exterior field of a rotating neutron star}},\ }\href
  {https://doi.org/10.1103/PhysRevD.61.081501} {\bibfield  {journal} {\bibinfo
  {journal} {Phys. Rev. D}\ }\textbf {\bibinfo {volume} {61}},\ \bibinfo
  {pages} {081501} (\bibinfo {year} {2000}{\natexlab{b}})},\ \Eprint
  {https://arxiv.org/abs/gr-qc/0001081} {arXiv:gr-qc/0001081} \BibitemShut
  {NoStop}%
\bibitem [{\citenamefont {Stute}\ and\ \citenamefont
  {Camenzind}(2002)}]{Stute:2002mqa}%
  \BibitemOpen
  \bibfield  {author} {\bibinfo {author} {\bibfnamefont {M.}~\bibnamefont
  {Stute}}\ and\ \bibinfo {author} {\bibfnamefont {M.}~\bibnamefont
  {Camenzind}},\ }\bibfield  {title} {\bibinfo {title} {{Towards a
  self-consistent relativistic model of the exterior gravitational field of
  rapidly rotating neutron stars}},\ }\href
  {https://doi.org/10.1046/j.1365-8711.2002.05820.x} {\bibfield  {journal}
  {\bibinfo  {journal} {Mon. Not. Roy. Astron. Soc.}\ }\textbf {\bibinfo
  {volume} {336}},\ \bibinfo {pages} {831} (\bibinfo {year} {2002})},\ \Eprint
  {https://arxiv.org/abs/astro-ph/0301466} {arXiv:astro-ph/0301466}
  \BibitemShut {NoStop}%
\bibitem [{\citenamefont {Pachon}\ \emph {et~al.}(2006)\citenamefont {Pachon},
  \citenamefont {Rueda},\ and\ \citenamefont {Sanabria-Gomez}}]{Pachon:2006an}%
  \BibitemOpen
  \bibfield  {author} {\bibinfo {author} {\bibfnamefont {L.~A.}\ \bibnamefont
  {Pachon}}, \bibinfo {author} {\bibfnamefont {J.~A.}\ \bibnamefont {Rueda}},\
  and\ \bibinfo {author} {\bibfnamefont {J.~D.}\ \bibnamefont
  {Sanabria-Gomez}},\ }\bibfield  {title} {\bibinfo {title} {{Realistic exact
  solution for the exterior field of a rotating neutron star}},\ }\href
  {https://doi.org/10.1103/PhysRevD.73.104038} {\bibfield  {journal} {\bibinfo
  {journal} {Phys. Rev. D}\ }\textbf {\bibinfo {volume} {73}},\ \bibinfo
  {pages} {104038} (\bibinfo {year} {2006})},\ \Eprint
  {https://arxiv.org/abs/gr-qc/0606060} {arXiv:gr-qc/0606060} \BibitemShut
  {NoStop}%
\bibitem [{\citenamefont {Teichmuller}\ \emph {et~al.}(2011)\citenamefont
  {Teichmuller}, \citenamefont {Fr\"ob},\ and\ \citenamefont
  {Maucher}}]{Teichmuller:2011px}%
  \BibitemOpen
  \bibfield  {author} {\bibinfo {author} {\bibfnamefont {C.}~\bibnamefont
  {Teichmuller}}, \bibinfo {author} {\bibfnamefont {M.~B.}\ \bibnamefont
  {Fr\"ob}},\ and\ \bibinfo {author} {\bibfnamefont {F.}~\bibnamefont
  {Maucher}},\ }\bibfield  {title} {\bibinfo {title} {{Analytical approximation
  of the exterior gravitational field of rotating neutron stars}},\ }\href
  {https://doi.org/10.1088/0264-9381/28/15/155015} {\bibfield  {journal}
  {\bibinfo  {journal} {Class. Quant. Grav.}\ }\textbf {\bibinfo {volume}
  {28}},\ \bibinfo {pages} {155015} (\bibinfo {year} {2011})},\ \Eprint
  {https://arxiv.org/abs/1102.5252} {arXiv:1102.5252 [gr-qc]} \BibitemShut
  {NoStop}%
\bibitem [{\citenamefont {Pappas}\ and\ \citenamefont
  {Apostolatos}(2013)}]{Pappas:2012nv}%
  \BibitemOpen
  \bibfield  {author} {\bibinfo {author} {\bibfnamefont {G.}~\bibnamefont
  {Pappas}}\ and\ \bibinfo {author} {\bibfnamefont {T.~A.}\ \bibnamefont
  {Apostolatos}},\ }\bibfield  {title} {\bibinfo {title} {{An all-purpose
  metric for the exterior of any kind of rotating neutron star}},\ }\href
  {https://doi.org/10.1093/mnras/sts556} {\bibfield  {journal} {\bibinfo
  {journal} {Mon. Not. Roy. Astron. Soc.}\ }\textbf {\bibinfo {volume} {429}},\
  \bibinfo {pages} {3007} (\bibinfo {year} {2013})},\ \Eprint
  {https://arxiv.org/abs/1209.6148} {arXiv:1209.6148 [gr-qc]} \BibitemShut
  {NoStop}%
\bibitem [{\citenamefont {Pappas}(2017)}]{Pappas:2016sye}%
  \BibitemOpen
  \bibfield  {author} {\bibinfo {author} {\bibfnamefont {G.}~\bibnamefont
  {Pappas}},\ }\bibfield  {title} {\bibinfo {title} {{An accurate metric for
  the spacetime around rotating neutron stars}},\ }\href
  {https://doi.org/10.1093/mnras/stx019} {\bibfield  {journal} {\bibinfo
  {journal} {Mon. Not. Roy. Astron. Soc.}\ }\textbf {\bibinfo {volume} {466}},\
  \bibinfo {pages} {4381} (\bibinfo {year} {2017})},\ \Eprint
  {https://arxiv.org/abs/1610.05370} {arXiv:1610.05370 [gr-qc]} \BibitemShut
  {NoStop}%
\bibitem [{\citenamefont {Stergioulas}\ and\ \citenamefont
  {Friedman}(1995)}]{Stergioulas:1994ea}%
  \BibitemOpen
  \bibfield  {author} {\bibinfo {author} {\bibfnamefont {N.}~\bibnamefont
  {Stergioulas}}\ and\ \bibinfo {author} {\bibfnamefont {J.~L.}\ \bibnamefont
  {Friedman}},\ }\bibfield  {title} {\bibinfo {title} {{Comparing models of
  rapidly rotating relativistic stars constructed by two numerical methods}},\
  }\href {https://doi.org/10.1086/175605} {\bibfield  {journal} {\bibinfo
  {journal} {Astrophys. J.}\ }\textbf {\bibinfo {volume} {444}},\ \bibinfo
  {pages} {306} (\bibinfo {year} {1995})},\ \Eprint
  {https://arxiv.org/abs/astro-ph/9411032} {arXiv:astro-ph/9411032}
  \BibitemShut {NoStop}%
\bibitem [{\citenamefont {Stergioulas}(2003)}]{Stergioulas:2003yp}%
  \BibitemOpen
  \bibfield  {author} {\bibinfo {author} {\bibfnamefont {N.}~\bibnamefont
  {Stergioulas}},\ }\bibfield  {title} {\bibinfo {title} {{Rotating Stars in
  Relativity}},\ }\href {https://doi.org/10.12942/lrr-2003-3} {\bibfield
  {journal} {\bibinfo  {journal} {Living Rev. Rel.}\ }\textbf {\bibinfo
  {volume} {6}},\ \bibinfo {pages} {3} (\bibinfo {year} {2003})},\ \Eprint
  {https://arxiv.org/abs/gr-qc/0302034} {arXiv:gr-qc/0302034} \BibitemShut
  {NoStop}%
\bibitem [{\citenamefont {{Hartle}}\ and\ \citenamefont
  {{Thorne}}(1968)}]{Hartle-Thorne2:1968}%
  \BibitemOpen
  \bibfield  {author} {\bibinfo {author} {\bibfnamefont {J.~B.}\ \bibnamefont
  {{Hartle}}}\ and\ \bibinfo {author} {\bibfnamefont {K.~S.}\ \bibnamefont
  {{Thorne}}},\ }\bibfield  {title} {\bibinfo {title} {{Slowly Rotating
  Relativistic Stars. II. Models for Neutron Stars and Supermassive Stars}},\
  }\href {https://doi.org/10.1086/149707} {\bibfield  {journal} {\bibinfo
  {journal} {ApJ}\ }\textbf {\bibinfo {volume} {153}},\ \bibinfo {pages} {807}
  (\bibinfo {year} {1968})}\BibitemShut {NoStop}%
\bibitem [{\citenamefont {{Hartle}}(1967)}]{Hartle-Thorne1:1967}%
  \BibitemOpen
  \bibfield  {author} {\bibinfo {author} {\bibfnamefont {J.~B.}\ \bibnamefont
  {{Hartle}}},\ }\bibfield  {title} {\bibinfo {title} {{Slowly Rotating
  Relativistic Stars. I. Equations of Structure}},\ }\href
  {https://doi.org/10.1086/149400} {\bibfield  {journal} {\bibinfo  {journal}
  {ApJ}\ }\textbf {\bibinfo {volume} {150}},\ \bibinfo {pages} {1005} (\bibinfo
  {year} {1967})}\BibitemShut {NoStop}%
\bibitem [{\citenamefont {{Hartle}}\ and\ \citenamefont
  {{Thorne}}(1969)}]{Hartle-Thorne3:1969}%
  \BibitemOpen
  \bibfield  {author} {\bibinfo {author} {\bibfnamefont {J.~B.}\ \bibnamefont
  {{Hartle}}}\ and\ \bibinfo {author} {\bibfnamefont {K.~S.}\ \bibnamefont
  {{Thorne}}},\ }\bibfield  {title} {\bibinfo {title} {{Slowly Rotating
  Relativistic Stars. III. Static Criterion for Stability}},\ }\href
  {https://doi.org/10.1086/150232} {\bibfield  {journal} {\bibinfo  {journal}
  {ApJ}\ }\textbf {\bibinfo {volume} {158}},\ \bibinfo {pages} {719} (\bibinfo
  {year} {1969})}\BibitemShut {NoStop}%
\bibitem [{\citenamefont {{Hartle}}(1970)}]{Hartle-Thorne4:1970}%
  \BibitemOpen
  \bibfield  {author} {\bibinfo {author} {\bibfnamefont {J.~B.}\ \bibnamefont
  {{Hartle}}},\ }\bibfield  {title} {\bibinfo {title} {{Slowly-Rotating
  Relativistic Stars.IV. Rotational Energy and Moment of Inertia for Stars in
  Differential Rotation}},\ }\href {https://doi.org/10.1086/150516} {\bibfield
  {journal} {\bibinfo  {journal} {ApJ}\ }\textbf {\bibinfo {volume} {161}},\
  \bibinfo {pages} {111} (\bibinfo {year} {1970})}\BibitemShut {NoStop}%
\bibitem [{\citenamefont {{Hartle}}\ \emph {et~al.}(1972)\citenamefont
  {{Hartle}}, \citenamefont {{Thorne}},\ and\ \citenamefont
  {{Chitre}}}]{Hartle-Thorne5:1972}%
  \BibitemOpen
  \bibfield  {author} {\bibinfo {author} {\bibfnamefont {J.~B.}\ \bibnamefont
  {{Hartle}}}, \bibinfo {author} {\bibfnamefont {K.~S.}\ \bibnamefont
  {{Thorne}}},\ and\ \bibinfo {author} {\bibfnamefont {S.~M.}\ \bibnamefont
  {{Chitre}}},\ }\bibfield  {title} {\bibinfo {title} {{Slowly Rotating
  Relativistic Stars.VI. Stability of the Quasiradial Modes}},\ }\href
  {https://doi.org/10.1086/151620} {\bibfield  {journal} {\bibinfo  {journal}
  {ApJ}\ }\textbf {\bibinfo {volume} {176}},\ \bibinfo {pages} {177} (\bibinfo
  {year} {1972})}\BibitemShut {NoStop}%
\bibitem [{\citenamefont {{Hartle}}(1975)}]{Hartle-Thorne6:1975}%
  \BibitemOpen
  \bibfield  {author} {\bibinfo {author} {\bibfnamefont {J.~B.}\ \bibnamefont
  {{Hartle}}},\ }\bibfield  {title} {\bibinfo {title} {{Slowly rotating
  relativistic stars. IIIA. The static stability criterion recovered.}},\
  }\href {https://doi.org/10.1086/153319} {\bibfield  {journal} {\bibinfo
  {journal} {ApJ}\ }\textbf {\bibinfo {volume} {195}},\ \bibinfo {pages} {203}
  (\bibinfo {year} {1975})}\BibitemShut {NoStop}%
\bibitem [{\citenamefont {Lense}\ and\ \citenamefont
  {Thirring}(1918)}]{Lense:1918}%
  \BibitemOpen
  \bibfield  {author} {\bibinfo {author} {\bibfnamefont {J.}~\bibnamefont
  {Lense}}\ and\ \bibinfo {author} {\bibfnamefont {H.}~\bibnamefont
  {Thirring}},\ }\bibfield  {title} {\bibinfo {title} {{On the Influence of the
  Proper Rotation of Central Bodies on the Motions of Planets and Moons
  According to Einstein’s Theory of Gravitation}},\ }\href
  {https://doi.org/1918PhyZ...19..156L} {\bibfield  {journal} {\bibinfo
  {journal} {Physikalische Zeitschrift}\ }\textbf {\bibinfo {volume} {19}},\
  \bibinfo {pages} {156} (\bibinfo {year} {1918})}\BibitemShut {NoStop}%
\bibitem [{\citenamefont {Vieira}\ \emph {et~al.}(2022)\citenamefont {Vieira},
  \citenamefont {Destounis},\ and\ \citenamefont {Kokkotas}}]{Vieira:2021ozg}%
  \BibitemOpen
  \bibfield  {author} {\bibinfo {author} {\bibfnamefont {H.~S.}\ \bibnamefont
  {Vieira}}, \bibinfo {author} {\bibfnamefont {K.}~\bibnamefont {Destounis}},\
  and\ \bibinfo {author} {\bibfnamefont {K.~D.}\ \bibnamefont {Kokkotas}},\
  }\bibfield  {title} {\bibinfo {title} {{Slowly-rotating curved acoustic black
  holes: Quasinormal modes, Hawking-Unruh radiation, and quasibound states}},\
  }\href {https://doi.org/10.1103/PhysRevD.105.045015} {\bibfield  {journal}
  {\bibinfo  {journal} {Phys. Rev. D}\ }\textbf {\bibinfo {volume} {105}},\
  \bibinfo {pages} {045015} (\bibinfo {year} {2022})},\ \Eprint
  {https://arxiv.org/abs/2112.08711} {arXiv:2112.08711 [gr-qc]} \BibitemShut
  {NoStop}%
\bibitem [{\citenamefont {Abramowicz}\ \emph {et~al.}(2003)\citenamefont
  {Abramowicz}, \citenamefont {Almergren}, \citenamefont {Kluzniak},\ and\
  \citenamefont {Thampan}}]{Abramowicz:2003rc}%
  \BibitemOpen
  \bibfield  {author} {\bibinfo {author} {\bibfnamefont {M.~A.}\ \bibnamefont
  {Abramowicz}}, \bibinfo {author} {\bibfnamefont {G.~J.~E.}\ \bibnamefont
  {Almergren}}, \bibinfo {author} {\bibfnamefont {W.}~\bibnamefont
  {Kluzniak}},\ and\ \bibinfo {author} {\bibfnamefont {A.~V.}\ \bibnamefont
  {Thampan}},\ }\bibfield  {title} {\bibinfo {title} {{Circular geodesics in
  the Hartle-Thorne metric}},\ }\href@noop {} {\  (\bibinfo {year} {2003})},\
  \Eprint {https://arxiv.org/abs/gr-qc/0312070} {arXiv:gr-qc/0312070}
  \BibitemShut {NoStop}%
\bibitem [{\citenamefont {{van der Klis}}\ \emph {et~al.}(1985)\citenamefont
  {{van der Klis}}, \citenamefont {{Jansen}}, \citenamefont {{van Paradijs}},
  \citenamefont {{Lewin}}, \citenamefont {{van den Heuvel}}, \citenamefont
  {{Trumper}},\ and\ \citenamefont
  {{Szatjno}}}]{vanderKlis:1985Natur.316..225V}%
  \BibitemOpen
  \bibfield  {author} {\bibinfo {author} {\bibfnamefont {M.}~\bibnamefont {{van
  der Klis}}}, \bibinfo {author} {\bibfnamefont {F.}~\bibnamefont {{Jansen}}},
  \bibinfo {author} {\bibfnamefont {J.}~\bibnamefont {{van Paradijs}}},
  \bibinfo {author} {\bibfnamefont {W.~H.~G.}\ \bibnamefont {{Lewin}}},
  \bibinfo {author} {\bibfnamefont {E.~P.~J.}\ \bibnamefont {{van den
  Heuvel}}}, \bibinfo {author} {\bibfnamefont {J.~E.}\ \bibnamefont
  {{Trumper}}},\ and\ \bibinfo {author} {\bibfnamefont {M.}~\bibnamefont
  {{Szatjno}}},\ }\bibfield  {title} {\bibinfo {title} {{Intensity-dependent
  quasi-periodic oscillations in the X-ray flux of GX5-1}},\ }\href
  {https://doi.org/10.1038/316225a0} {\bibfield  {journal} {\bibinfo  {journal}
  {Nature}\ }\textbf {\bibinfo {volume} {316}},\ \bibinfo {pages} {225}
  (\bibinfo {year} {1985})}\BibitemShut {NoStop}%
\bibitem [{\citenamefont {Abramowicz}\ \emph {et~al.}(2002)\citenamefont
  {Abramowicz}, \citenamefont {Almergren}, \citenamefont {Kluzniak},
  \citenamefont {Thampan},\ and\ \citenamefont
  {Wallinder}}]{Abramowicz:2002di}%
  \BibitemOpen
  \bibfield  {author} {\bibinfo {author} {\bibfnamefont {M.~A.}\ \bibnamefont
  {Abramowicz}}, \bibinfo {author} {\bibfnamefont {G.~J.~E.}\ \bibnamefont
  {Almergren}}, \bibinfo {author} {\bibfnamefont {W.}~\bibnamefont {Kluzniak}},
  \bibinfo {author} {\bibfnamefont {A.~V.}\ \bibnamefont {Thampan}},\ and\
  \bibinfo {author} {\bibfnamefont {F.}~\bibnamefont {Wallinder}},\ }\bibfield
  {title} {\bibinfo {title} {{Holonomy invariance, orbital resonances, and
  kilohertz QPO(s)}},\ }\href {https://doi.org/10.1088/0264-9381/19/8/103}
  {\bibfield  {journal} {\bibinfo  {journal} {Class. Quant. Grav.}\ }\textbf
  {\bibinfo {volume} {19}},\ \bibinfo {pages} {L57} (\bibinfo {year} {2002})},\
  \Eprint {https://arxiv.org/abs/gr-qc/0202020} {arXiv:gr-qc/0202020}
  \BibitemShut {NoStop}%
\bibitem [{\citenamefont {Ingram}\ \emph {et~al.}(2016)\citenamefont {Ingram},
  \citenamefont {van~der Klis}, \citenamefont {Middleton}, \citenamefont
  {Done}, \citenamefont {Altamirano}, \citenamefont {Heil}, \citenamefont
  {Uttley},\ and\ \citenamefont {Axelsson}}]{Ingram:2016tbq}%
  \BibitemOpen
  \bibfield  {author} {\bibinfo {author} {\bibfnamefont {A.}~\bibnamefont
  {Ingram}}, \bibinfo {author} {\bibfnamefont {M.}~\bibnamefont {van~der
  Klis}}, \bibinfo {author} {\bibfnamefont {M.}~\bibnamefont {Middleton}},
  \bibinfo {author} {\bibfnamefont {C.}~\bibnamefont {Done}}, \bibinfo {author}
  {\bibfnamefont {D.}~\bibnamefont {Altamirano}}, \bibinfo {author}
  {\bibfnamefont {L.}~\bibnamefont {Heil}}, \bibinfo {author} {\bibfnamefont
  {P.}~\bibnamefont {Uttley}},\ and\ \bibinfo {author} {\bibfnamefont
  {M.}~\bibnamefont {Axelsson}},\ }\bibfield  {title} {\bibinfo {title} {{A
  quasi-periodic modulation of the iron line centroid energy in the black hole
  binary H1743\ensuremath{-}322}},\ }\href
  {https://doi.org/10.1093/mnras/stw1245} {\bibfield  {journal} {\bibinfo
  {journal} {Mon. Not. Roy. Astron. Soc.}\ }\textbf {\bibinfo {volume} {461}},\
  \bibinfo {pages} {1967} (\bibinfo {year} {2016})},\ \Eprint
  {https://arxiv.org/abs/1607.02866} {arXiv:1607.02866 [astro-ph.HE]}
  \BibitemShut {NoStop}%
\bibitem [{\citenamefont {Urbancov\'a}\ \emph {et~al.}(2019)\citenamefont
  {Urbancov\'a}, \citenamefont {Urbanec}, \citenamefont {T\"or\"ok},
  \citenamefont {Stuchl\'\i{}k}, \citenamefont {Blaschke},\ and\ \citenamefont
  {Miller}}]{Urbancova:2019btk}%
  \BibitemOpen
  \bibfield  {author} {\bibinfo {author} {\bibfnamefont {G.}~\bibnamefont
  {Urbancov\'a}}, \bibinfo {author} {\bibfnamefont {M.}~\bibnamefont
  {Urbanec}}, \bibinfo {author} {\bibfnamefont {G.}~\bibnamefont {T\"or\"ok}},
  \bibinfo {author} {\bibfnamefont {Z.}~\bibnamefont {Stuchl\'\i{}k}}, \bibinfo
  {author} {\bibfnamefont {M.}~\bibnamefont {Blaschke}},\ and\ \bibinfo
  {author} {\bibfnamefont {J.~C.}\ \bibnamefont {Miller}},\ }\bibfield  {title}
  {\bibinfo {title} {{Epicyclic Oscillations in the Hartle\textendash{}Thorne
  External Geometry}},\ }\href {https://doi.org/10.3847/1538-4357/ab1b4c}
  {\bibfield  {journal} {\bibinfo  {journal} {Astrophys. J.}\ }\textbf
  {\bibinfo {volume} {877}},\ \bibinfo {pages} {66} (\bibinfo {year} {2019})},\
  \Eprint {https://arxiv.org/abs/1905.00730} {arXiv:1905.00730 [astro-ph.HE]}
  \BibitemShut {NoStop}%
\bibitem [{\citenamefont {Sulieva}\ \emph {et~al.}(2022)\citenamefont
  {Sulieva}, \citenamefont {Boshkayev}, \citenamefont {Nurbakyt}, \citenamefont
  {Quevedo}, \citenamefont {Taukenova}, \citenamefont {Tlemissov},
  \citenamefont {Tlemissova},\ and\ \citenamefont
  {Urazalina}}]{Sulieva:2022kpc}%
  \BibitemOpen
  \bibfield  {author} {\bibinfo {author} {\bibfnamefont {G.}~\bibnamefont
  {Sulieva}}, \bibinfo {author} {\bibfnamefont {K.}~\bibnamefont {Boshkayev}},
  \bibinfo {author} {\bibfnamefont {G.}~\bibnamefont {Nurbakyt}}, \bibinfo
  {author} {\bibfnamefont {H.}~\bibnamefont {Quevedo}}, \bibinfo {author}
  {\bibfnamefont {A.}~\bibnamefont {Taukenova}}, \bibinfo {author}
  {\bibfnamefont {A.}~\bibnamefont {Tlemissov}}, \bibinfo {author}
  {\bibfnamefont {Z.}~\bibnamefont {Tlemissova}},\ and\ \bibinfo {author}
  {\bibfnamefont {A.}~\bibnamefont {Urazalina}},\ }\bibfield  {title} {\bibinfo
  {title} {{Adiabatic theory of motion of bodies in the Hartle-Thorne
  spacetime}},\ }\href@noop {} {\  (\bibinfo {year} {2022})},\ \Eprint
  {https://arxiv.org/abs/2205.04217} {arXiv:2205.04217 [gr-qc]} \BibitemShut
  {NoStop}%
\bibitem [{\citenamefont {Stella}(2001)}]{Stella:2000gd}%
  \BibitemOpen
  \bibfield  {author} {\bibinfo {author} {\bibfnamefont {L.}~\bibnamefont
  {Stella}},\ }\bibfield  {title} {\bibinfo {title} {{The relativistic
  precession model for qpos in low mass x-ray binaries}},\ }\href
  {https://doi.org/10.1063/1.1434649} {\bibfield  {journal} {\bibinfo
  {journal} {AIP Conf. Proc.}\ }\textbf {\bibinfo {volume} {599}},\ \bibinfo
  {pages} {365} (\bibinfo {year} {2001})},\ \Eprint
  {https://arxiv.org/abs/astro-ph/0011395} {arXiv:astro-ph/0011395}
  \BibitemShut {NoStop}%
\bibitem [{\citenamefont {Rezzolla}\ \emph {et~al.}(2003)\citenamefont
  {Rezzolla}, \citenamefont {Yoshida}, \citenamefont {Maccarone},\ and\
  \citenamefont {Zanotti}}]{Rezzolla:2003zx}%
  \BibitemOpen
  \bibfield  {author} {\bibinfo {author} {\bibfnamefont {L.}~\bibnamefont
  {Rezzolla}}, \bibinfo {author} {\bibfnamefont {S.}~\bibnamefont {Yoshida}},
  \bibinfo {author} {\bibfnamefont {T.~J.}\ \bibnamefont {Maccarone}},\ and\
  \bibinfo {author} {\bibfnamefont {O.}~\bibnamefont {Zanotti}},\ }\bibfield
  {title} {\bibinfo {title} {{A New simple model for high frequency quasi
  periodic oscillations in black hole candidates}},\ }\href
  {https://doi.org/10.1046/j.1365-8711.2003.07018.x} {\bibfield  {journal}
  {\bibinfo  {journal} {Mon. Not. Roy. Astron. Soc.}\ }\textbf {\bibinfo
  {volume} {344}},\ \bibinfo {pages} {L37} (\bibinfo {year} {2003})},\ \Eprint
  {https://arxiv.org/abs/astro-ph/0307487} {arXiv:astro-ph/0307487}
  \BibitemShut {NoStop}%
\bibitem [{\citenamefont {Carter}(1968)}]{Carter:1968rr}%
  \BibitemOpen
  \bibfield  {author} {\bibinfo {author} {\bibfnamefont {B.}~\bibnamefont
  {Carter}},\ }\bibfield  {title} {\bibinfo {title} {{Global structure of the
  Kerr family of gravitational fields}},\ }\href
  {https://doi.org/10.1103/PhysRev.174.1559} {\bibfield  {journal} {\bibinfo
  {journal} {Phys. Rev.}\ }\textbf {\bibinfo {volume} {174}},\ \bibinfo {pages}
  {1559} (\bibinfo {year} {1968})}\BibitemShut {NoStop}%
\bibitem [{\citenamefont {Johannsen}\ and\ \citenamefont
  {Psaltis}(2011)}]{Johannsen:2011dh}%
  \BibitemOpen
  \bibfield  {author} {\bibinfo {author} {\bibfnamefont {T.}~\bibnamefont
  {Johannsen}}\ and\ \bibinfo {author} {\bibfnamefont {D.}~\bibnamefont
  {Psaltis}},\ }\bibfield  {title} {\bibinfo {title} {{A Metric for Rapidly
  Spinning Black Holes Suitable for Strong-Field Tests of the No-Hair
  Theorem}},\ }\href {https://doi.org/10.1103/PhysRevD.83.124015} {\bibfield
  {journal} {\bibinfo  {journal} {Phys. Rev. D}\ }\textbf {\bibinfo {volume}
  {83}},\ \bibinfo {pages} {124015} (\bibinfo {year} {2011})},\ \Eprint
  {https://arxiv.org/abs/1105.3191} {arXiv:1105.3191 [gr-qc]} \BibitemShut
  {NoStop}%
\bibitem [{\citenamefont {Papadopoulos}\ and\ \citenamefont
  {Kokkotas}(2018)}]{Papadopoulos:2018nvd}%
  \BibitemOpen
  \bibfield  {author} {\bibinfo {author} {\bibfnamefont {G.~O.}\ \bibnamefont
  {Papadopoulos}}\ and\ \bibinfo {author} {\bibfnamefont {K.~D.}\ \bibnamefont
  {Kokkotas}},\ }\bibfield  {title} {\bibinfo {title} {{Preserving Kerr
  symmetries in deformed spacetimes}},\ }\href
  {https://doi.org/10.1088/1361-6382/aad7f4} {\bibfield  {journal} {\bibinfo
  {journal} {Class. Quant. Grav.}\ }\textbf {\bibinfo {volume} {35}},\ \bibinfo
  {pages} {185014} (\bibinfo {year} {2018})},\ \Eprint
  {https://arxiv.org/abs/1807.08594} {arXiv:1807.08594 [gr-qc]} \BibitemShut
  {NoStop}%
\bibitem [{\citenamefont {Konoplya}\ \emph {et~al.}(2018)\citenamefont
  {Konoplya}, \citenamefont {Stuchl\'\i{}k},\ and\ \citenamefont
  {Zhidenko}}]{Konoplya:2018arm}%
  \BibitemOpen
  \bibfield  {author} {\bibinfo {author} {\bibfnamefont {R.~A.}\ \bibnamefont
  {Konoplya}}, \bibinfo {author} {\bibfnamefont {Z.}~\bibnamefont
  {Stuchl\'\i{}k}},\ and\ \bibinfo {author} {\bibfnamefont {A.}~\bibnamefont
  {Zhidenko}},\ }\bibfield  {title} {\bibinfo {title} {{Axisymmetric black
  holes allowing for separation of variables in the Klein-Gordon and
  Hamilton-Jacobi equations}},\ }\href
  {https://doi.org/10.1103/PhysRevD.97.084044} {\bibfield  {journal} {\bibinfo
  {journal} {Phys. Rev. D}\ }\textbf {\bibinfo {volume} {97}},\ \bibinfo
  {pages} {084044} (\bibinfo {year} {2018})},\ \Eprint
  {https://arxiv.org/abs/1801.07195} {arXiv:1801.07195 [gr-qc]} \BibitemShut
  {NoStop}%
\bibitem [{\citenamefont {Papadopoulos}\ and\ \citenamefont
  {Kokkotas}(2021)}]{Papadopoulos:2020kxu}%
  \BibitemOpen
  \bibfield  {author} {\bibinfo {author} {\bibfnamefont {G.~O.}\ \bibnamefont
  {Papadopoulos}}\ and\ \bibinfo {author} {\bibfnamefont {K.~D.}\ \bibnamefont
  {Kokkotas}},\ }\bibfield  {title} {\bibinfo {title} {{On Kerr black hole
  deformations admitting a Carter constant and an invariant criterion for the
  separability of the wave equation}},\ }\href
  {https://doi.org/10.1007/s10714-021-02795-2} {\bibfield  {journal} {\bibinfo
  {journal} {Gen. Rel. Grav.}\ }\textbf {\bibinfo {volume} {53}},\ \bibinfo
  {pages} {21} (\bibinfo {year} {2021})},\ \Eprint
  {https://arxiv.org/abs/2007.12125} {arXiv:2007.12125 [gr-qc]} \BibitemShut
  {NoStop}%
\bibitem [{\citenamefont {Konoplya}\ and\ \citenamefont
  {Zhidenko}(2021)}]{Konoplya:2021slg}%
  \BibitemOpen
  \bibfield  {author} {\bibinfo {author} {\bibfnamefont {R.~A.}\ \bibnamefont
  {Konoplya}}\ and\ \bibinfo {author} {\bibfnamefont {A.}~\bibnamefont
  {Zhidenko}},\ }\bibfield  {title} {\bibinfo {title} {{Shadows of parametrized
  axially symmetric black holes allowing for separation of variables}},\ }\href
  {https://doi.org/10.1103/PhysRevD.103.104033} {\bibfield  {journal} {\bibinfo
   {journal} {Phys. Rev. D}\ }\textbf {\bibinfo {volume} {103}},\ \bibinfo
  {pages} {104033} (\bibinfo {year} {2021})},\ \Eprint
  {https://arxiv.org/abs/2103.03855} {arXiv:2103.03855 [gr-qc]} \BibitemShut
  {NoStop}%
\bibitem [{\citenamefont {Glampedakis}\ and\ \citenamefont
  {Babak}(2006)}]{Glampedakis:2005cf}%
  \BibitemOpen
  \bibfield  {author} {\bibinfo {author} {\bibfnamefont {K.}~\bibnamefont
  {Glampedakis}}\ and\ \bibinfo {author} {\bibfnamefont {S.}~\bibnamefont
  {Babak}},\ }\bibfield  {title} {\bibinfo {title} {{Mapping spacetimes with
  LISA: Inspiral of a test-body in a `quasi-Kerr' field}},\ }\href
  {https://doi.org/10.1088/0264-9381/23/12/013} {\bibfield  {journal} {\bibinfo
   {journal} {Class. Quant. Grav.}\ }\textbf {\bibinfo {volume} {23}},\
  \bibinfo {pages} {4167} (\bibinfo {year} {2006})},\ \Eprint
  {https://arxiv.org/abs/gr-qc/0510057} {arXiv:gr-qc/0510057} \BibitemShut
  {NoStop}%
\bibitem [{\citenamefont {Cornish}(2001)}]{Cornish:2001jy}%
  \BibitemOpen
  \bibfield  {author} {\bibinfo {author} {\bibfnamefont {N.~J.}\ \bibnamefont
  {Cornish}},\ }\bibfield  {title} {\bibinfo {title} {{Chaos and gravitational
  waves}},\ }\href {https://doi.org/10.1103/PhysRevD.64.084011} {\bibfield
  {journal} {\bibinfo  {journal} {Phys. Rev. D}\ }\textbf {\bibinfo {volume}
  {64}},\ \bibinfo {pages} {084011} (\bibinfo {year} {2001})},\ \Eprint
  {https://arxiv.org/abs/gr-qc/0106062} {arXiv:gr-qc/0106062} \BibitemShut
  {NoStop}%
\bibitem [{\citenamefont {Cornish}\ and\ \citenamefont
  {Levin}(2003)}]{Cornish:2003ig}%
  \BibitemOpen
  \bibfield  {author} {\bibinfo {author} {\bibfnamefont {N.~J.}\ \bibnamefont
  {Cornish}}\ and\ \bibinfo {author} {\bibfnamefont {J.~J.}\ \bibnamefont
  {Levin}},\ }\bibfield  {title} {\bibinfo {title} {{Lyapunov timescales and
  black hole binaries}},\ }\href {https://doi.org/10.1088/0264-9381/20/9/304}
  {\bibfield  {journal} {\bibinfo  {journal} {Class. Quant. Grav.}\ }\textbf
  {\bibinfo {volume} {20}},\ \bibinfo {pages} {1649} (\bibinfo {year}
  {2003})},\ \Eprint {https://arxiv.org/abs/gr-qc/0304056}
  {arXiv:gr-qc/0304056} \BibitemShut {NoStop}%
\bibitem [{\citenamefont {Verhaaren}\ and\ \citenamefont
  {Hirschmann}(2010)}]{Verhaaren:2009md}%
  \BibitemOpen
  \bibfield  {author} {\bibinfo {author} {\bibfnamefont {C.}~\bibnamefont
  {Verhaaren}}\ and\ \bibinfo {author} {\bibfnamefont {E.~W.}\ \bibnamefont
  {Hirschmann}},\ }\bibfield  {title} {\bibinfo {title} {{Chaotic orbits for
  spinning particles in Schwarzschild spacetime}},\ }\href
  {https://doi.org/10.1103/PhysRevD.81.124034} {\bibfield  {journal} {\bibinfo
  {journal} {Phys. Rev. D}\ }\textbf {\bibinfo {volume} {81}},\ \bibinfo
  {pages} {124034} (\bibinfo {year} {2010})},\ \Eprint
  {https://arxiv.org/abs/0912.0031} {arXiv:0912.0031 [gr-qc]} \BibitemShut
  {NoStop}%
\bibitem [{\citenamefont {Barausse}\ \emph {et~al.}(2007)\citenamefont
  {Barausse}, \citenamefont {Rezzolla}, \citenamefont {Petroff},\ and\
  \citenamefont {Ansorg}}]{Barausse:2006vt}%
  \BibitemOpen
  \bibfield  {author} {\bibinfo {author} {\bibfnamefont {E.}~\bibnamefont
  {Barausse}}, \bibinfo {author} {\bibfnamefont {L.}~\bibnamefont {Rezzolla}},
  \bibinfo {author} {\bibfnamefont {D.}~\bibnamefont {Petroff}},\ and\ \bibinfo
  {author} {\bibfnamefont {M.}~\bibnamefont {Ansorg}},\ }\bibfield  {title}
  {\bibinfo {title} {{Gravitational waves from Extreme Mass Ratio Inspirals in
  non-pure Kerr spacetimes}},\ }\href
  {https://doi.org/10.1103/PhysRevD.75.064026} {\bibfield  {journal} {\bibinfo
  {journal} {Phys. Rev. D}\ }\textbf {\bibinfo {volume} {75}},\ \bibinfo
  {pages} {064026} (\bibinfo {year} {2007})},\ \Eprint
  {https://arxiv.org/abs/gr-qc/0612123} {arXiv:gr-qc/0612123} \BibitemShut
  {NoStop}%
\bibitem [{\citenamefont {Barausse}\ and\ \citenamefont
  {Rezzolla}(2008)}]{Barausse:2007dy}%
  \BibitemOpen
  \bibfield  {author} {\bibinfo {author} {\bibfnamefont {E.}~\bibnamefont
  {Barausse}}\ and\ \bibinfo {author} {\bibfnamefont {L.}~\bibnamefont
  {Rezzolla}},\ }\bibfield  {title} {\bibinfo {title} {{The Influence of the
  hydrodynamic drag from an accretion torus on extreme mass-ratio inspirals}},\
  }\href {https://doi.org/10.1103/PhysRevD.77.104027} {\bibfield  {journal}
  {\bibinfo  {journal} {Phys. Rev. D}\ }\textbf {\bibinfo {volume} {77}},\
  \bibinfo {pages} {104027} (\bibinfo {year} {2008})},\ \Eprint
  {https://arxiv.org/abs/0711.4558} {arXiv:0711.4558 [gr-qc]} \BibitemShut
  {NoStop}%
\bibitem [{\citenamefont {Suzuki}\ and\ \citenamefont
  {Maeda}(1997)}]{Suzuki:1996gm}%
  \BibitemOpen
  \bibfield  {author} {\bibinfo {author} {\bibfnamefont {S.}~\bibnamefont
  {Suzuki}}\ and\ \bibinfo {author} {\bibfnamefont {K.-i.}\ \bibnamefont
  {Maeda}},\ }\bibfield  {title} {\bibinfo {title} {{Chaos in Schwarzschild
  space-time: The motion of a spinning particle}},\ }\href
  {https://doi.org/10.1103/PhysRevD.55.4848} {\bibfield  {journal} {\bibinfo
  {journal} {Phys. Rev. D}\ }\textbf {\bibinfo {volume} {55}},\ \bibinfo
  {pages} {4848} (\bibinfo {year} {1997})},\ \Eprint
  {https://arxiv.org/abs/gr-qc/9604020} {arXiv:gr-qc/9604020} \BibitemShut
  {NoStop}%
\bibitem [{\citenamefont {Zelenka}\ \emph {et~al.}(2020)\citenamefont
  {Zelenka}, \citenamefont {Lukes-Gerakopoulos}, \citenamefont {Witzany},\ and\
  \citenamefont {Kop\'a\v{c}ek}}]{Zelenka:2019nyp}%
  \BibitemOpen
  \bibfield  {author} {\bibinfo {author} {\bibfnamefont {O.}~\bibnamefont
  {Zelenka}}, \bibinfo {author} {\bibfnamefont {G.}~\bibnamefont
  {Lukes-Gerakopoulos}}, \bibinfo {author} {\bibfnamefont {V.}~\bibnamefont
  {Witzany}},\ and\ \bibinfo {author} {\bibfnamefont {O.}~\bibnamefont
  {Kop\'a\v{c}ek}},\ }\bibfield  {title} {\bibinfo {title} {{Growth of
  resonances and chaos for a spinning test particle in the Schwarzschild
  background}},\ }\href {https://doi.org/10.1103/PhysRevD.101.024037}
  {\bibfield  {journal} {\bibinfo  {journal} {Phys. Rev. D}\ }\textbf {\bibinfo
  {volume} {101}},\ \bibinfo {pages} {024037} (\bibinfo {year} {2020})},\
  \Eprint {https://arxiv.org/abs/1911.00414} {arXiv:1911.00414 [gr-qc]}
  \BibitemShut {NoStop}%
\bibitem [{\citenamefont {Lukes-Gerakopoulos}\ \emph
  {et~al.}(2014)\citenamefont {Lukes-Gerakopoulos}, \citenamefont
  {Contopoulos},\ and\ \citenamefont
  {Apostolatos}}]{Lukes-Gerakopoulos:2014dpa}%
  \BibitemOpen
  \bibfield  {author} {\bibinfo {author} {\bibfnamefont {G.}~\bibnamefont
  {Lukes-Gerakopoulos}}, \bibinfo {author} {\bibfnamefont {G.}~\bibnamefont
  {Contopoulos}},\ and\ \bibinfo {author} {\bibfnamefont {T.~A.}\ \bibnamefont
  {Apostolatos}},\ }\bibfield  {title} {\bibinfo {title} {{Non-Linear Effects
  in Non-Kerr spacetimes}},\ }\href
  {https://doi.org/10.1007/978-3-319-06761-2_16} {\bibfield  {journal}
  {\bibinfo  {journal} {Springer Proc. Phys.}\ }\textbf {\bibinfo {volume}
  {157}},\ \bibinfo {pages} {129} (\bibinfo {year} {2014})},\ \Eprint
  {https://arxiv.org/abs/1408.4697} {arXiv:1408.4697 [gr-qc]} \BibitemShut
  {NoStop}%
\bibitem [{\citenamefont {Stuchl\'\i{}k}\ \emph {et~al.}(2020)\citenamefont
  {Stuchl\'\i{}k}, \citenamefont {Kolo\v{s}}, \citenamefont {Kov\'a\v{r}},
  \citenamefont {Slan\'y},\ and\ \citenamefont {Tursunov}}]{Stuchlik:2020rls}%
  \BibitemOpen
  \bibfield  {author} {\bibinfo {author} {\bibfnamefont {Z.}~\bibnamefont
  {Stuchl\'\i{}k}}, \bibinfo {author} {\bibfnamefont {M.}~\bibnamefont
  {Kolo\v{s}}}, \bibinfo {author} {\bibfnamefont {J.}~\bibnamefont
  {Kov\'a\v{r}}}, \bibinfo {author} {\bibfnamefont {P.}~\bibnamefont
  {Slan\'y}},\ and\ \bibinfo {author} {\bibfnamefont {A.}~\bibnamefont
  {Tursunov}},\ }\bibfield  {title} {\bibinfo {title} {{Influence of Cosmic
  Repulsion and Magnetic Fields on Accretion Disks Rotating around Kerr Black
  Holes}},\ }\href {https://doi.org/10.3390/universe6020026} {\bibfield
  {journal} {\bibinfo  {journal} {Universe}\ }\textbf {\bibinfo {volume} {6}},\
  \bibinfo {pages} {26} (\bibinfo {year} {2020})}\BibitemShut {NoStop}%
\bibitem [{\citenamefont {Stuchl\'\i{}k}\ and\ \citenamefont
  {Vrba}(2021)}]{Stuchlik:2021gwg}%
  \BibitemOpen
  \bibfield  {author} {\bibinfo {author} {\bibfnamefont {Z.}~\bibnamefont
  {Stuchl\'\i{}k}}\ and\ \bibinfo {author} {\bibfnamefont {J.}~\bibnamefont
  {Vrba}},\ }\bibfield  {title} {\bibinfo {title} {{Supermassive black holes
  surrounded by dark matter modeled as anisotropic fluid: epicyclic
  oscillations and their fitting to observed QPOs}},\ }\href
  {https://doi.org/10.1088/1475-7516/2021/11/059} {\bibfield  {journal}
  {\bibinfo  {journal} {JCAP}\ }\textbf {\bibinfo {volume} {11}}\bibfield
  {number} {\bibinfo  {number} { (11)},\ \bibinfo {pages} {059}},\ }\Eprint
  {https://arxiv.org/abs/2110.07411} {arXiv:2110.07411 [gr-qc]} \BibitemShut
  {NoStop}%
\bibitem [{\citenamefont {Leung}\ \emph {et~al.}(2022)\citenamefont {Leung},
  \citenamefont {Yip}, \citenamefont {Cheong},\ and\ \citenamefont
  {Li}}]{Leung:2022mvm}%
  \BibitemOpen
  \bibfield  {author} {\bibinfo {author} {\bibfnamefont {M.~Y.}\ \bibnamefont
  {Leung}}, \bibinfo {author} {\bibfnamefont {A.~K.~L.}\ \bibnamefont {Yip}},
  \bibinfo {author} {\bibfnamefont {P.~C.-K.}\ \bibnamefont {Cheong}},\ and\
  \bibinfo {author} {\bibfnamefont {T.~G.~F.}\ \bibnamefont {Li}},\ }\bibfield
  {title} {\bibinfo {title} {{Oscillations of highly magnetized non-rotating
  neutron stars}},\ }\href {https://doi.org/10.1038/s42005-022-01112-w}
  {\bibfield  {journal} {\bibinfo  {journal} {Commun. Phys.}\ }\textbf
  {\bibinfo {volume} {5}},\ \bibinfo {pages} {334} (\bibinfo {year} {2022})},\
  \Eprint {https://arxiv.org/abs/2303.05684} {arXiv:2303.05684 [astro-ph.HE]}
  \BibitemShut {NoStop}%
\bibitem [{\citenamefont {Apostolatos}\ \emph {et~al.}(2009)\citenamefont
  {Apostolatos}, \citenamefont {Lukes-Gerakopoulos},\ and\ \citenamefont
  {Contopoulos}}]{Apostolatos:2009vu}%
  \BibitemOpen
  \bibfield  {author} {\bibinfo {author} {\bibfnamefont {T.~A.}\ \bibnamefont
  {Apostolatos}}, \bibinfo {author} {\bibfnamefont {G.}~\bibnamefont
  {Lukes-Gerakopoulos}},\ and\ \bibinfo {author} {\bibfnamefont
  {G.}~\bibnamefont {Contopoulos}},\ }\bibfield  {title} {\bibinfo {title}
  {{How to Observe a Non-Kerr Spacetime Using Gravitational Waves}},\ }\href
  {https://doi.org/10.1103/PhysRevLett.103.111101} {\bibfield  {journal}
  {\bibinfo  {journal} {Phys. Rev. Lett.}\ }\textbf {\bibinfo {volume} {103}},\
  \bibinfo {pages} {111101} (\bibinfo {year} {2009})},\ \Eprint
  {https://arxiv.org/abs/0906.0093} {arXiv:0906.0093 [gr-qc]} \BibitemShut
  {NoStop}%
\bibitem [{\citenamefont {Lukes-Gerakopoulos}\ \emph
  {et~al.}(2010)\citenamefont {Lukes-Gerakopoulos}, \citenamefont
  {Apostolatos},\ and\ \citenamefont
  {Contopoulos}}]{Lukes-Gerakopoulos:2010ipp}%
  \BibitemOpen
  \bibfield  {author} {\bibinfo {author} {\bibfnamefont {G.}~\bibnamefont
  {Lukes-Gerakopoulos}}, \bibinfo {author} {\bibfnamefont {T.~A.}\ \bibnamefont
  {Apostolatos}},\ and\ \bibinfo {author} {\bibfnamefont {G.}~\bibnamefont
  {Contopoulos}},\ }\bibfield  {title} {\bibinfo {title} {{Observable signature
  of a background deviating from the Kerr metric}},\ }\href
  {https://doi.org/10.1103/PhysRevD.81.124005} {\bibfield  {journal} {\bibinfo
  {journal} {Phys. Rev. D}\ }\textbf {\bibinfo {volume} {81}},\ \bibinfo
  {pages} {124005} (\bibinfo {year} {2010})},\ \Eprint
  {https://arxiv.org/abs/1003.3120} {arXiv:1003.3120 [gr-qc]} \BibitemShut
  {NoStop}%
\bibitem [{\citenamefont {Destounis}\ \emph {et~al.}(2020)\citenamefont
  {Destounis}, \citenamefont {Suvorov},\ and\ \citenamefont
  {Kokkotas}}]{Destounis:2020kss}%
  \BibitemOpen
  \bibfield  {author} {\bibinfo {author} {\bibfnamefont {K.}~\bibnamefont
  {Destounis}}, \bibinfo {author} {\bibfnamefont {A.~G.}\ \bibnamefont
  {Suvorov}},\ and\ \bibinfo {author} {\bibfnamefont {K.~D.}\ \bibnamefont
  {Kokkotas}},\ }\bibfield  {title} {\bibinfo {title} {{Testing spacetime
  symmetry through gravitational waves from extreme-mass-ratio inspirals}},\
  }\href {https://doi.org/10.1103/PhysRevD.102.064041} {\bibfield  {journal}
  {\bibinfo  {journal} {Phys. Rev. D}\ }\textbf {\bibinfo {volume} {102}},\
  \bibinfo {pages} {064041} (\bibinfo {year} {2020})},\ \Eprint
  {https://arxiv.org/abs/2009.00028} {arXiv:2009.00028 [gr-qc]} \BibitemShut
  {NoStop}%
\bibitem [{\citenamefont {Chen}\ \emph {et~al.}(2022)\citenamefont {Chen},
  \citenamefont {Lin},\ and\ \citenamefont {Patel}}]{Chen:2022znf}%
  \BibitemOpen
  \bibfield  {author} {\bibinfo {author} {\bibfnamefont {C.-Y.}\ \bibnamefont
  {Chen}}, \bibinfo {author} {\bibfnamefont {F.-L.}\ \bibnamefont {Lin}},\ and\
  \bibinfo {author} {\bibfnamefont {A.}~\bibnamefont {Patel}},\ }\bibfield
  {title} {\bibinfo {title} {{Resonant islands of effective-one-body
  dynamics}},\ }\href {https://doi.org/10.1103/PhysRevD.106.084064} {\bibfield
  {journal} {\bibinfo  {journal} {Phys. Rev. D}\ }\textbf {\bibinfo {volume}
  {106}},\ \bibinfo {pages} {084064} (\bibinfo {year} {2022})},\ \Eprint
  {https://arxiv.org/abs/2206.10966} {arXiv:2206.10966 [gr-qc]} \BibitemShut
  {NoStop}%
\bibitem [{\citenamefont {Chen}\ \emph {et~al.}(2023)\citenamefont {Chen},
  \citenamefont {Chiang},\ and\ \citenamefont {Patel}}]{Chen:2023gwm}%
  \BibitemOpen
  \bibfield  {author} {\bibinfo {author} {\bibfnamefont {C.-Y.}\ \bibnamefont
  {Chen}}, \bibinfo {author} {\bibfnamefont {H.-W.}\ \bibnamefont {Chiang}},\
  and\ \bibinfo {author} {\bibfnamefont {A.}~\bibnamefont {Patel}},\ }\bibfield
   {title} {\bibinfo {title} {{Resonant orbits of rotating black holes beyond
  circularity: Discontinuity along a parameter shift}},\ }\href
  {https://doi.org/10.1103/PhysRevD.108.064016} {\bibfield  {journal} {\bibinfo
   {journal} {Phys. Rev. D}\ }\textbf {\bibinfo {volume} {108}},\ \bibinfo
  {pages} {064016} (\bibinfo {year} {2023})},\ \Eprint
  {https://arxiv.org/abs/2306.08356} {arXiv:2306.08356 [gr-qc]} \BibitemShut
  {NoStop}%
\bibitem [{\citenamefont {Deich}\ \emph {et~al.}(2022)\citenamefont {Deich},
  \citenamefont {C\'ardenas-Avenda\~no},\ and\ \citenamefont
  {Yunes}}]{Deich:2022vna}%
  \BibitemOpen
  \bibfield  {author} {\bibinfo {author} {\bibfnamefont {A.}~\bibnamefont
  {Deich}}, \bibinfo {author} {\bibfnamefont {A.}~\bibnamefont
  {C\'ardenas-Avenda\~no}},\ and\ \bibinfo {author} {\bibfnamefont
  {N.}~\bibnamefont {Yunes}},\ }\bibfield  {title} {\bibinfo {title} {{Chaos in
  quadratic gravity}},\ }\href {https://doi.org/10.1103/PhysRevD.106.024040}
  {\bibfield  {journal} {\bibinfo  {journal} {Phys. Rev. D}\ }\textbf {\bibinfo
  {volume} {106}},\ \bibinfo {pages} {024040} (\bibinfo {year} {2022})},\
  \Eprint {https://arxiv.org/abs/2203.00524} {arXiv:2203.00524 [gr-qc]}
  \BibitemShut {NoStop}%
\bibitem [{\citenamefont {Destounis}\ \emph {et~al.}(2021)\citenamefont
  {Destounis}, \citenamefont {Suvorov},\ and\ \citenamefont
  {Kokkotas}}]{Destounis:2021mqv}%
  \BibitemOpen
  \bibfield  {author} {\bibinfo {author} {\bibfnamefont {K.}~\bibnamefont
  {Destounis}}, \bibinfo {author} {\bibfnamefont {A.~G.}\ \bibnamefont
  {Suvorov}},\ and\ \bibinfo {author} {\bibfnamefont {K.~D.}\ \bibnamefont
  {Kokkotas}},\ }\bibfield  {title} {\bibinfo {title} {{Gravitational-wave
  glitches in chaotic extreme-mass-ratio inspirals}},\ }\href
  {https://doi.org/10.1103/PhysRevLett.126.141102} {\bibfield  {journal}
  {\bibinfo  {journal} {Phys. Rev. Lett.}\ }\textbf {\bibinfo {volume} {126}},\
  \bibinfo {pages} {141102} (\bibinfo {year} {2021})},\ \Eprint
  {https://arxiv.org/abs/2103.05643} {arXiv:2103.05643 [gr-qc]} \BibitemShut
  {NoStop}%
\bibitem [{\citenamefont {Destounis}\ and\ \citenamefont
  {Kokkotas}(2021)}]{Destounis:2021rko}%
  \BibitemOpen
  \bibfield  {author} {\bibinfo {author} {\bibfnamefont {K.}~\bibnamefont
  {Destounis}}\ and\ \bibinfo {author} {\bibfnamefont {K.~D.}\ \bibnamefont
  {Kokkotas}},\ }\bibfield  {title} {\bibinfo {title} {{Gravitational-wave
  glitches: Resonant islands and frequency jumps in nonintegrable
  extreme-mass-ratio inspirals}},\ }\href
  {https://doi.org/10.1103/PhysRevD.104.064023} {\bibfield  {journal} {\bibinfo
   {journal} {Phys. Rev. D}\ }\textbf {\bibinfo {volume} {104}},\ \bibinfo
  {pages} {064023} (\bibinfo {year} {2021})},\ \Eprint
  {https://arxiv.org/abs/2108.02782} {arXiv:2108.02782 [gr-qc]} \BibitemShut
  {NoStop}%
\bibitem [{\citenamefont {Destounis}\ \emph
  {et~al.}(2023{\natexlab{b}})\citenamefont {Destounis}, \citenamefont {Huez},\
  and\ \citenamefont {Kokkotas}}]{Destounis:2023gpw}%
  \BibitemOpen
  \bibfield  {author} {\bibinfo {author} {\bibfnamefont {K.}~\bibnamefont
  {Destounis}}, \bibinfo {author} {\bibfnamefont {G.}~\bibnamefont {Huez}},\
  and\ \bibinfo {author} {\bibfnamefont {K.~D.}\ \bibnamefont {Kokkotas}},\
  }\bibfield  {title} {\bibinfo {title} {{Geodesics and gravitational waves in
  chaotic extreme-mass-ratio inspirals: The curious case of Zipoy-Voorhees
  black-hole mimickers}},\ }\href@noop {} {\  (\bibinfo {year}
  {2023}{\natexlab{b}})},\ \Eprint {https://arxiv.org/abs/2301.11483}
  {arXiv:2301.11483 [gr-qc]} \BibitemShut {NoStop}%
\bibitem [{\citenamefont {Destounis}\ \emph
  {et~al.}(2023{\natexlab{c}})\citenamefont {Destounis}, \citenamefont
  {Angeloni}, \citenamefont {Vaglio},\ and\ \citenamefont
  {Pani}}]{Destounis:2023khj}%
  \BibitemOpen
  \bibfield  {author} {\bibinfo {author} {\bibfnamefont {K.}~\bibnamefont
  {Destounis}}, \bibinfo {author} {\bibfnamefont {F.}~\bibnamefont {Angeloni}},
  \bibinfo {author} {\bibfnamefont {M.}~\bibnamefont {Vaglio}},\ and\ \bibinfo
  {author} {\bibfnamefont {P.}~\bibnamefont {Pani}},\ }\bibfield  {title}
  {\bibinfo {title} {{Extreme-mass-ratio inspirals into rotating boson stars:
  nonintegrability, chaos, and transient resonances}},\ }\href@noop {} {\
  (\bibinfo {year} {2023}{\natexlab{c}})},\ \Eprint
  {https://arxiv.org/abs/2305.05691} {arXiv:2305.05691 [gr-qc]} \BibitemShut
  {NoStop}%
\bibitem [{\citenamefont {Amaro-Seoane}\ \emph {et~al.}(2017)\citenamefont
  {Amaro-Seoane} \emph {et~al.}}]{LISA:2017pwj}%
  \BibitemOpen
  \bibfield  {author} {\bibinfo {author} {\bibfnamefont {P.}~\bibnamefont
  {Amaro-Seoane}} \emph {et~al.} (\bibinfo {collaboration} {LISA}),\ }\bibfield
   {title} {\bibinfo {title} {{Laser Interferometer Space Antenna}},\
  }\href@noop {} {\  (\bibinfo {year} {2017})},\ \Eprint
  {https://arxiv.org/abs/1702.00786} {arXiv:1702.00786 [astro-ph.IM]}
  \BibitemShut {NoStop}%
\bibitem [{\citenamefont {Barausse}\ \emph {et~al.}(2020)\citenamefont
  {Barausse} \emph {et~al.}}]{Barausse:2020rsu}%
  \BibitemOpen
  \bibfield  {author} {\bibinfo {author} {\bibfnamefont {E.}~\bibnamefont
  {Barausse}} \emph {et~al.},\ }\bibfield  {title} {\bibinfo {title}
  {{Prospects for Fundamental Physics with LISA}},\ }\href
  {https://doi.org/10.1007/s10714-020-02691-1} {\bibfield  {journal} {\bibinfo
  {journal} {Gen. Rel. Grav.}\ }\textbf {\bibinfo {volume} {52}},\ \bibinfo
  {pages} {81} (\bibinfo {year} {2020})},\ \Eprint
  {https://arxiv.org/abs/2001.09793} {arXiv:2001.09793 [gr-qc]} \BibitemShut
  {NoStop}%
\bibitem [{\citenamefont {Seoane}\ \emph {et~al.}(2023)\citenamefont {Seoane}
  \emph {et~al.}}]{LISA:2022yao}%
  \BibitemOpen
  \bibfield  {author} {\bibinfo {author} {\bibfnamefont {P.~A.}\ \bibnamefont
  {Seoane}} \emph {et~al.} (\bibinfo {collaboration} {LISA}),\ }\bibfield
  {title} {\bibinfo {title} {{Astrophysics with the Laser Interferometer Space
  Antenna}},\ }\href {https://doi.org/10.1007/s41114-022-00041-y} {\bibfield
  {journal} {\bibinfo  {journal} {Living Rev. Rel.}\ }\textbf {\bibinfo
  {volume} {26}},\ \bibinfo {pages} {2} (\bibinfo {year} {2023})},\ \Eprint
  {https://arxiv.org/abs/2203.06016} {arXiv:2203.06016 [gr-qc]} \BibitemShut
  {NoStop}%
\bibitem [{\citenamefont {Arun}\ \emph {et~al.}(2022)\citenamefont {Arun} \emph
  {et~al.}}]{LISA:2022kgy}%
  \BibitemOpen
  \bibfield  {author} {\bibinfo {author} {\bibfnamefont {K.~G.}\ \bibnamefont
  {Arun}} \emph {et~al.} (\bibinfo {collaboration} {LISA}),\ }\bibfield
  {title} {\bibinfo {title} {{New horizons for fundamental physics with
  LISA}},\ }\href {https://doi.org/10.1007/s41114-022-00036-9} {\bibfield
  {journal} {\bibinfo  {journal} {Living Rev. Rel.}\ }\textbf {\bibinfo
  {volume} {25}},\ \bibinfo {pages} {4} (\bibinfo {year} {2022})},\ \Eprint
  {https://arxiv.org/abs/2205.01597} {arXiv:2205.01597 [gr-qc]} \BibitemShut
  {NoStop}%
\bibitem [{\citenamefont {Karnesis}\ \emph {et~al.}(2022)\citenamefont
  {Karnesis} \emph {et~al.}}]{Karnesis:2022vdp}%
  \BibitemOpen
  \bibfield  {author} {\bibinfo {author} {\bibfnamefont {N.}~\bibnamefont
  {Karnesis}} \emph {et~al.},\ }\bibfield  {title} {\bibinfo {title} {{The
  Laser Interferometer Space Antenna mission in Greece White Paper}},\
  }\href@noop {} {\  (\bibinfo {year} {2022})},\ \Eprint
  {https://arxiv.org/abs/2209.04358} {arXiv:2209.04358 [gr-qc]} \BibitemShut
  {NoStop}%
\bibitem [{\citenamefont {Kostaros}\ and\ \citenamefont
  {Pappas}(2022)}]{Kostaros:2021usv}%
  \BibitemOpen
  \bibfield  {author} {\bibinfo {author} {\bibfnamefont {K.}~\bibnamefont
  {Kostaros}}\ and\ \bibinfo {author} {\bibfnamefont {G.}~\bibnamefont
  {Pappas}},\ }\bibfield  {title} {\bibinfo {title} {{Chaotic photon orbits and
  shadows of a non-Kerr object described by the Hartle\textendash{}Thorne
  spacetime}},\ }\href {https://doi.org/10.1088/1361-6382/ac7028} {\bibfield
  {journal} {\bibinfo  {journal} {Class. Quant. Grav.}\ }\textbf {\bibinfo
  {volume} {39}},\ \bibinfo {pages} {134001} (\bibinfo {year} {2022})},\
  \Eprint {https://arxiv.org/abs/2111.09367} {arXiv:2111.09367 [gr-qc]}
  \BibitemShut {NoStop}%
\bibitem [{\citenamefont {Contopoulos}(2003)}]{Contopoulos_book}%
  \BibitemOpen
  \bibfield  {author} {\bibinfo {author} {\bibfnamefont {G.}~\bibnamefont
  {Contopoulos}},\ }\href {https://doi.org/10.1063/1.1634536} {\emph {\bibinfo
  {title} {Order and Chaos in Dynamical Astronomy}}}\ (\bibinfo  {publisher}
  {Springer-Verlag},\ \bibinfo {address} {New York},\ \bibinfo {year}
  {2003})\BibitemShut {NoStop}%
\bibitem [{\citenamefont {Glampedakis}(2005)}]{Glampedakis:2005hs}%
  \BibitemOpen
  \bibfield  {author} {\bibinfo {author} {\bibfnamefont {K.}~\bibnamefont
  {Glampedakis}},\ }\bibfield  {title} {\bibinfo {title} {{Extreme mass ratio
  inspirals: LISA's unique probe of black hole gravity}},\ }\href
  {https://doi.org/10.1088/0264-9381/22/15/004} {\bibfield  {journal} {\bibinfo
   {journal} {Class. Quant. Grav.}\ }\textbf {\bibinfo {volume} {22}},\
  \bibinfo {pages} {S605} (\bibinfo {year} {2005})},\ \Eprint
  {https://arxiv.org/abs/gr-qc/0509024} {arXiv:gr-qc/0509024} \BibitemShut
  {NoStop}%
\bibitem [{\citenamefont {Amaro-Seoane}(2018)}]{Amaro-Seoane:2012lgq}%
  \BibitemOpen
  \bibfield  {author} {\bibinfo {author} {\bibfnamefont {P.}~\bibnamefont
  {Amaro-Seoane}},\ }\bibfield  {title} {\bibinfo {title} {{Relativistic
  dynamics and extreme mass ratio inspirals}},\ }\href
  {https://doi.org/10.1007/s41114-018-0013-8} {\bibfield  {journal} {\bibinfo
  {journal} {Living Rev. Rel.}\ }\textbf {\bibinfo {volume} {21}},\ \bibinfo
  {pages} {4} (\bibinfo {year} {2018})},\ \Eprint
  {https://arxiv.org/abs/1205.5240} {arXiv:1205.5240 [astro-ph.CO]}
  \BibitemShut {NoStop}%
\bibitem [{\citenamefont
  {Lukes-Gerakopoulos}(2012)}]{Lukes-Gerakopoulos:2012qpc}%
  \BibitemOpen
  \bibfield  {author} {\bibinfo {author} {\bibfnamefont {G.}~\bibnamefont
  {Lukes-Gerakopoulos}},\ }\bibfield  {title} {\bibinfo {title} {{The
  non-integrability of the Zipoy-Voorhees metric}},\ }\href
  {https://doi.org/10.1103/PhysRevD.86.044013} {\bibfield  {journal} {\bibinfo
  {journal} {Phys. Rev. D}\ }\textbf {\bibinfo {volume} {86}},\ \bibinfo
  {pages} {044013} (\bibinfo {year} {2012})},\ \Eprint
  {https://arxiv.org/abs/1206.0660} {arXiv:1206.0660 [gr-qc]} \BibitemShut
  {NoStop}%
\bibitem [{\citenamefont {Glampedakis}\ and\ \citenamefont
  {Pappas}(2018)}]{Glampedakis:2017cgd}%
  \BibitemOpen
  \bibfield  {author} {\bibinfo {author} {\bibfnamefont {K.}~\bibnamefont
  {Glampedakis}}\ and\ \bibinfo {author} {\bibfnamefont {G.}~\bibnamefont
  {Pappas}},\ }\bibfield  {title} {\bibinfo {title} {{How well can ultracompact
  bodies imitate black hole ringdowns?}},\ }\href
  {https://doi.org/10.1103/PhysRevD.97.041502} {\bibfield  {journal} {\bibinfo
  {journal} {Phys. Rev. D}\ }\textbf {\bibinfo {volume} {97}},\ \bibinfo
  {pages} {041502} (\bibinfo {year} {2018})},\ \Eprint
  {https://arxiv.org/abs/1710.02136} {arXiv:1710.02136 [gr-qc]} \BibitemShut
  {NoStop}%
\bibitem [{\citenamefont {Yagi}\ \emph {et~al.}(2014)\citenamefont {Yagi},
  \citenamefont {Kyutoku}, \citenamefont {Pappas}, \citenamefont {Yunes},\ and\
  \citenamefont {Apostolatos}}]{Yagi:2014bxa}%
  \BibitemOpen
  \bibfield  {author} {\bibinfo {author} {\bibfnamefont {K.}~\bibnamefont
  {Yagi}}, \bibinfo {author} {\bibfnamefont {K.}~\bibnamefont {Kyutoku}},
  \bibinfo {author} {\bibfnamefont {G.}~\bibnamefont {Pappas}}, \bibinfo
  {author} {\bibfnamefont {N.}~\bibnamefont {Yunes}},\ and\ \bibinfo {author}
  {\bibfnamefont {T.~A.}\ \bibnamefont {Apostolatos}},\ }\bibfield  {title}
  {\bibinfo {title} {{Effective No-Hair Relations for Neutron Stars and Quark
  Stars: Relativistic Results}},\ }\href
  {https://doi.org/10.1103/PhysRevD.89.124013} {\bibfield  {journal} {\bibinfo
  {journal} {Phys. Rev. D}\ }\textbf {\bibinfo {volume} {89}},\ \bibinfo
  {pages} {124013} (\bibinfo {year} {2014})},\ \Eprint
  {https://arxiv.org/abs/1403.6243} {arXiv:1403.6243 [gr-qc]} \BibitemShut
  {NoStop}%
\bibitem [{\citenamefont {Raposo}\ \emph {et~al.}(2019)\citenamefont {Raposo},
  \citenamefont {Pani},\ and\ \citenamefont {Emparan}}]{Raposo:2018xkf}%
  \BibitemOpen
  \bibfield  {author} {\bibinfo {author} {\bibfnamefont {G.}~\bibnamefont
  {Raposo}}, \bibinfo {author} {\bibfnamefont {P.}~\bibnamefont {Pani}},\ and\
  \bibinfo {author} {\bibfnamefont {R.}~\bibnamefont {Emparan}},\ }\bibfield
  {title} {\bibinfo {title} {{Exotic compact objects with soft hair}},\ }\href
  {https://doi.org/10.1103/PhysRevD.99.104050} {\bibfield  {journal} {\bibinfo
  {journal} {Phys. Rev. D}\ }\textbf {\bibinfo {volume} {99}},\ \bibinfo
  {pages} {104050} (\bibinfo {year} {2019})},\ \Eprint
  {https://arxiv.org/abs/1812.07615} {arXiv:1812.07615 [gr-qc]} \BibitemShut
  {NoStop}%
\bibitem [{\citenamefont {Pacilio}\ \emph {et~al.}(2020)\citenamefont
  {Pacilio}, \citenamefont {Vaglio}, \citenamefont {Maselli},\ and\
  \citenamefont {Pani}}]{Pacilio:2020jza}%
  \BibitemOpen
  \bibfield  {author} {\bibinfo {author} {\bibfnamefont {C.}~\bibnamefont
  {Pacilio}}, \bibinfo {author} {\bibfnamefont {M.}~\bibnamefont {Vaglio}},
  \bibinfo {author} {\bibfnamefont {A.}~\bibnamefont {Maselli}},\ and\ \bibinfo
  {author} {\bibfnamefont {P.}~\bibnamefont {Pani}},\ }\bibfield  {title}
  {\bibinfo {title} {{Gravitational-wave detectors as particle-physics
  laboratories: Constraining scalar interactions with a coherent inspiral model
  of boson-star binaries}},\ }\href
  {https://doi.org/10.1103/PhysRevD.102.083002} {\bibfield  {journal} {\bibinfo
   {journal} {Phys. Rev. D}\ }\textbf {\bibinfo {volume} {102}},\ \bibinfo
  {pages} {083002} (\bibinfo {year} {2020})},\ \Eprint
  {https://arxiv.org/abs/2007.05264} {arXiv:2007.05264 [gr-qc]} \BibitemShut
  {NoStop}%
\bibitem [{\citenamefont {Vaglio}\ \emph {et~al.}(2022)\citenamefont {Vaglio},
  \citenamefont {Pacilio}, \citenamefont {Maselli},\ and\ \citenamefont
  {Pani}}]{Vaglio:2022flq}%
  \BibitemOpen
  \bibfield  {author} {\bibinfo {author} {\bibfnamefont {M.}~\bibnamefont
  {Vaglio}}, \bibinfo {author} {\bibfnamefont {C.}~\bibnamefont {Pacilio}},
  \bibinfo {author} {\bibfnamefont {A.}~\bibnamefont {Maselli}},\ and\ \bibinfo
  {author} {\bibfnamefont {P.}~\bibnamefont {Pani}},\ }\bibfield  {title}
  {\bibinfo {title} {{Multipolar structure of rotating boson stars}},\ }\href
  {https://doi.org/10.1103/PhysRevD.105.124020} {\bibfield  {journal} {\bibinfo
   {journal} {Phys. Rev. D}\ }\textbf {\bibinfo {volume} {105}},\ \bibinfo
  {pages} {124020} (\bibinfo {year} {2022})},\ \Eprint
  {https://arxiv.org/abs/2203.07442} {arXiv:2203.07442 [gr-qc]} \BibitemShut
  {NoStop}%
\bibitem [{\citenamefont {Vaglio}\ \emph {et~al.}(2023)\citenamefont {Vaglio},
  \citenamefont {Pacilio}, \citenamefont {Maselli},\ and\ \citenamefont
  {Pani}}]{Vaglio:2023lrd}%
  \BibitemOpen
  \bibfield  {author} {\bibinfo {author} {\bibfnamefont {M.}~\bibnamefont
  {Vaglio}}, \bibinfo {author} {\bibfnamefont {C.}~\bibnamefont {Pacilio}},
  \bibinfo {author} {\bibfnamefont {A.}~\bibnamefont {Maselli}},\ and\ \bibinfo
  {author} {\bibfnamefont {P.}~\bibnamefont {Pani}},\ }\bibfield  {title}
  {\bibinfo {title} {{Bayesian parameter estimation on boson-star binary
  signals with a coherent inspiral template and spin-dependent quadrupolar
  corrections}},\ }\href@noop {} {\  (\bibinfo {year} {2023})},\ \Eprint
  {https://arxiv.org/abs/2302.13954} {arXiv:2302.13954 [gr-qc]} \BibitemShut
  {NoStop}%
\bibitem [{\citenamefont {Mignemi}\ and\ \citenamefont
  {Stewart}(1993)}]{Mignemi:1992nt}%
  \BibitemOpen
  \bibfield  {author} {\bibinfo {author} {\bibfnamefont {S.}~\bibnamefont
  {Mignemi}}\ and\ \bibinfo {author} {\bibfnamefont {N.~R.}\ \bibnamefont
  {Stewart}},\ }\bibfield  {title} {\bibinfo {title} {{Charged black holes in
  effective string theory}},\ }\href {https://doi.org/10.1103/PhysRevD.47.5259}
  {\bibfield  {journal} {\bibinfo  {journal} {Phys. Rev. D}\ }\textbf {\bibinfo
  {volume} {47}},\ \bibinfo {pages} {5259} (\bibinfo {year} {1993})},\ \Eprint
  {https://arxiv.org/abs/hep-th/9212146} {arXiv:hep-th/9212146} \BibitemShut
  {NoStop}%
\bibitem [{\citenamefont {Mignemi}(1995)}]{Mignemi:1993ce}%
  \BibitemOpen
  \bibfield  {author} {\bibinfo {author} {\bibfnamefont {S.}~\bibnamefont
  {Mignemi}},\ }\bibfield  {title} {\bibinfo {title} {{Dyonic black holes in
  effective string theory}},\ }\href {https://doi.org/10.1103/PhysRevD.51.934}
  {\bibfield  {journal} {\bibinfo  {journal} {Phys. Rev. D}\ }\textbf {\bibinfo
  {volume} {51}},\ \bibinfo {pages} {934} (\bibinfo {year} {1995})},\ \Eprint
  {https://arxiv.org/abs/hep-th/9303102} {arXiv:hep-th/9303102} \BibitemShut
  {NoStop}%
\bibitem [{\citenamefont {Canizares}\ \emph {et~al.}(2012)\citenamefont
  {Canizares}, \citenamefont {Gair},\ and\ \citenamefont
  {Sopuerta}}]{Canizares:2012is}%
  \BibitemOpen
  \bibfield  {author} {\bibinfo {author} {\bibfnamefont {P.}~\bibnamefont
  {Canizares}}, \bibinfo {author} {\bibfnamefont {J.~R.}\ \bibnamefont
  {Gair}},\ and\ \bibinfo {author} {\bibfnamefont {C.~F.}\ \bibnamefont
  {Sopuerta}},\ }\bibfield  {title} {\bibinfo {title} {{Testing Chern-Simons
  Modified Gravity with Gravitational-Wave Detections of Extreme-Mass-Ratio
  Binaries}},\ }\href {https://doi.org/10.1103/PhysRevD.86.044010} {\bibfield
  {journal} {\bibinfo  {journal} {Phys. Rev. D}\ }\textbf {\bibinfo {volume}
  {86}},\ \bibinfo {pages} {044010} (\bibinfo {year} {2012})},\ \Eprint
  {https://arxiv.org/abs/1205.1253} {arXiv:1205.1253 [gr-qc]} \BibitemShut
  {NoStop}%
\bibitem [{\citenamefont {De~Falco}\ \emph {et~al.}(2020)\citenamefont
  {De~Falco}, \citenamefont {Bakala},\ and\ \citenamefont
  {Falanga}}]{DeFalco:2020vli}%
  \BibitemOpen
  \bibfield  {author} {\bibinfo {author} {\bibfnamefont {V.}~\bibnamefont
  {De~Falco}}, \bibinfo {author} {\bibfnamefont {P.}~\bibnamefont {Bakala}},\
  and\ \bibinfo {author} {\bibfnamefont {M.}~\bibnamefont {Falanga}},\
  }\bibfield  {title} {\bibinfo {title} {{Three-dimensional general
  relativistic Poynting-Robertson effect. III. Static and nonspherical
  quadrupolar massive source}},\ }\href
  {https://doi.org/10.1103/PhysRevD.101.124031} {\bibfield  {journal} {\bibinfo
   {journal} {Phys. Rev. D}\ }\textbf {\bibinfo {volume} {101}},\ \bibinfo
  {pages} {124031} (\bibinfo {year} {2020})},\ \Eprint
  {https://arxiv.org/abs/2006.01452} {arXiv:2006.01452 [gr-qc]} \BibitemShut
  {NoStop}%
\bibitem [{\citenamefont {De~Falco}\ and\ \citenamefont
  {Wielgus}(2021)}]{DeFalco:2021nfk}%
  \BibitemOpen
  \bibfield  {author} {\bibinfo {author} {\bibfnamefont {V.}~\bibnamefont
  {De~Falco}}\ and\ \bibinfo {author} {\bibfnamefont {M.}~\bibnamefont
  {Wielgus}},\ }\bibfield  {title} {\bibinfo {title} {{Three-dimensional
  general relativistic Poynting-Robertson effect. IV. Slowly rotating and
  nonspherical quadrupolar massive source}},\ }\href
  {https://doi.org/10.1103/PhysRevD.103.084056} {\bibfield  {journal} {\bibinfo
   {journal} {Phys. Rev. D}\ }\textbf {\bibinfo {volume} {103}},\ \bibinfo
  {pages} {084056} (\bibinfo {year} {2021})},\ \Eprint
  {https://arxiv.org/abs/2103.17165} {arXiv:2103.17165 [gr-qc]} \BibitemShut
  {NoStop}%
\bibitem [{\citenamefont {Cunha}\ \emph {et~al.}(2016)\citenamefont {Cunha},
  \citenamefont {Grover}, \citenamefont {Herdeiro}, \citenamefont {Radu},
  \citenamefont {Runarsson},\ and\ \citenamefont {Wittig}}]{Cunha:2016bjh}%
  \BibitemOpen
  \bibfield  {author} {\bibinfo {author} {\bibfnamefont {P.~V.~P.}\
  \bibnamefont {Cunha}}, \bibinfo {author} {\bibfnamefont {J.}~\bibnamefont
  {Grover}}, \bibinfo {author} {\bibfnamefont {C.}~\bibnamefont {Herdeiro}},
  \bibinfo {author} {\bibfnamefont {E.}~\bibnamefont {Radu}}, \bibinfo {author}
  {\bibfnamefont {H.}~\bibnamefont {Runarsson}},\ and\ \bibinfo {author}
  {\bibfnamefont {A.}~\bibnamefont {Wittig}},\ }\bibfield  {title} {\bibinfo
  {title} {{Chaotic lensing around boson stars and Kerr black holes with scalar
  hair}},\ }\href {https://doi.org/10.1103/PhysRevD.94.104023} {\bibfield
  {journal} {\bibinfo  {journal} {Phys. Rev. D}\ }\textbf {\bibinfo {volume}
  {94}},\ \bibinfo {pages} {104023} (\bibinfo {year} {2016})},\ \Eprint
  {https://arxiv.org/abs/1609.01340} {arXiv:1609.01340 [gr-qc]} \BibitemShut
  {NoStop}%
\bibitem [{\citenamefont {Shipley}\ and\ \citenamefont
  {Dolan}(2016)}]{Shipley:2016omi}%
  \BibitemOpen
  \bibfield  {author} {\bibinfo {author} {\bibfnamefont {J.}~\bibnamefont
  {Shipley}}\ and\ \bibinfo {author} {\bibfnamefont {S.~R.}\ \bibnamefont
  {Dolan}},\ }\bibfield  {title} {\bibinfo {title} {{Binary black hole shadows,
  chaotic scattering and the Cantor set}},\ }\href
  {https://doi.org/10.1088/0264-9381/33/17/175001} {\bibfield  {journal}
  {\bibinfo  {journal} {Class. Quant. Grav.}\ }\textbf {\bibinfo {volume}
  {33}},\ \bibinfo {pages} {175001} (\bibinfo {year} {2016})},\ \Eprint
  {https://arxiv.org/abs/1603.04469} {arXiv:1603.04469 [gr-qc]} \BibitemShut
  {NoStop}%
\bibitem [{\citenamefont {Benhar}\ \emph {et~al.}(2005)\citenamefont {Benhar},
  \citenamefont {Ferrari}, \citenamefont {Gualtieri},\ and\ \citenamefont
  {Marassi}}]{Benhar:2005gi}%
  \BibitemOpen
  \bibfield  {author} {\bibinfo {author} {\bibfnamefont {O.}~\bibnamefont
  {Benhar}}, \bibinfo {author} {\bibfnamefont {V.}~\bibnamefont {Ferrari}},
  \bibinfo {author} {\bibfnamefont {L.}~\bibnamefont {Gualtieri}},\ and\
  \bibinfo {author} {\bibfnamefont {S.}~\bibnamefont {Marassi}},\ }\bibfield
  {title} {\bibinfo {title} {{Perturbative approach to the structure of rapidly
  rotating neutron stars}},\ }\href
  {https://doi.org/10.1103/PhysRevD.72.044028} {\bibfield  {journal} {\bibinfo
  {journal} {Phys. Rev. D}\ }\textbf {\bibinfo {volume} {72}},\ \bibinfo
  {pages} {044028} (\bibinfo {year} {2005})},\ \Eprint
  {https://arxiv.org/abs/gr-qc/0504068} {arXiv:gr-qc/0504068} \BibitemShut
  {NoStop}%
\bibitem [{\citenamefont {Glampedakis}\ and\ \citenamefont
  {Pappas}(2019)}]{Glampedakis:2018blj}%
  \BibitemOpen
  \bibfield  {author} {\bibinfo {author} {\bibfnamefont {K.}~\bibnamefont
  {Glampedakis}}\ and\ \bibinfo {author} {\bibfnamefont {G.}~\bibnamefont
  {Pappas}},\ }\bibfield  {title} {\bibinfo {title} {{Modification of photon
  trapping orbits as a diagnostic of non-Kerr spacetimes}},\ }\href
  {https://doi.org/10.1103/PhysRevD.99.124041} {\bibfield  {journal} {\bibinfo
  {journal} {Phys. Rev. D}\ }\textbf {\bibinfo {volume} {99}},\ \bibinfo
  {pages} {124041} (\bibinfo {year} {2019})},\ \Eprint
  {https://arxiv.org/abs/1806.09333} {arXiv:1806.09333 [gr-qc]} \BibitemShut
  {NoStop}%
\bibitem [{\citenamefont {Cardoso}\ and\ \citenamefont
  {Pani}(2019)}]{Cardoso:2019rvt}%
  \BibitemOpen
  \bibfield  {author} {\bibinfo {author} {\bibfnamefont {V.}~\bibnamefont
  {Cardoso}}\ and\ \bibinfo {author} {\bibfnamefont {P.}~\bibnamefont {Pani}},\
  }\bibfield  {title} {\bibinfo {title} {{Testing the nature of dark compact
  objects: a status report}},\ }\href
  {https://doi.org/10.1007/s41114-019-0020-4} {\bibfield  {journal} {\bibinfo
  {journal} {Living Rev. Rel.}\ }\textbf {\bibinfo {volume} {22}},\ \bibinfo
  {pages} {4} (\bibinfo {year} {2019})},\ \Eprint
  {https://arxiv.org/abs/1904.05363} {arXiv:1904.05363 [gr-qc]} \BibitemShut
  {NoStop}%
\bibitem [{\citenamefont {Maggio}\ \emph {et~al.}(2020)\citenamefont {Maggio},
  \citenamefont {Pani},\ and\ \citenamefont {Raposo}}]{Maggio2020}%
  \BibitemOpen
  \bibfield  {author} {\bibinfo {author} {\bibfnamefont {E.}~\bibnamefont
  {Maggio}}, \bibinfo {author} {\bibfnamefont {P.}~\bibnamefont {Pani}},\ and\
  \bibinfo {author} {\bibfnamefont {G.}~\bibnamefont {Raposo}},\ }\bibinfo
  {title} {Testing the nature of dark compact objects with gravitational
  waves},\ in\ \href {https://doi.org/10.1007/978-981-15-4702-7_29-1} {\emph
  {\bibinfo {booktitle} {Handbook of Gravitational Wave Astronomy}}},\ \bibinfo
  {editor} {edited by\ \bibinfo {editor} {\bibfnamefont {C.}~\bibnamefont
  {Bambi}}, \bibinfo {editor} {\bibfnamefont {S.}~\bibnamefont {Katsanevas}},\
  and\ \bibinfo {editor} {\bibfnamefont {K.~D.}\ \bibnamefont {Kokkotas}}}\
  (\bibinfo  {publisher} {Springer Singapore},\ \bibinfo {address}
  {Singapore},\ \bibinfo {year} {2020})\ pp.\ \bibinfo {pages}
  {1--37}\BibitemShut {NoStop}%
\bibitem [{\citenamefont {Bini}\ \emph {et~al.}(2013)\citenamefont {Bini},
  \citenamefont {Boshkayev}, \citenamefont {Ruffini},\ and\ \citenamefont
  {Siutsou}}]{Bini:2013tia}%
  \BibitemOpen
  \bibfield  {author} {\bibinfo {author} {\bibfnamefont {D.}~\bibnamefont
  {Bini}}, \bibinfo {author} {\bibfnamefont {K.}~\bibnamefont {Boshkayev}},
  \bibinfo {author} {\bibfnamefont {R.}~\bibnamefont {Ruffini}},\ and\ \bibinfo
  {author} {\bibfnamefont {I.}~\bibnamefont {Siutsou}},\ }\bibfield  {title}
  {\bibinfo {title} {{Equatorial Circular Geodesics in the Hartle-Thorne
  Spacetime}},\ }\href {https://doi.org/10.1393/ncc/i2013-11483-8} {\bibfield
  {journal} {\bibinfo  {journal} {Nuovo Cim. C}\ }\textbf {\bibinfo {volume}
  {036}},\ \bibinfo {pages} {31} (\bibinfo {year} {2013})},\ \Eprint
  {https://arxiv.org/abs/1306.4792} {arXiv:1306.4792 [gr-qc]} \BibitemShut
  {NoStop}%
\bibitem [{\citenamefont {Contopoulos}\ \emph {et~al.}(2011)\citenamefont
  {Contopoulos}, \citenamefont {Lukes-Gerakopoulos},\ and\ \citenamefont
  {Apostolatos}}]{Contopoulos:2011dz}%
  \BibitemOpen
  \bibfield  {author} {\bibinfo {author} {\bibfnamefont {G.}~\bibnamefont
  {Contopoulos}}, \bibinfo {author} {\bibfnamefont {G.}~\bibnamefont
  {Lukes-Gerakopoulos}},\ and\ \bibinfo {author} {\bibfnamefont {T.~A.}\
  \bibnamefont {Apostolatos}},\ }\bibfield  {title} {\bibinfo {title} {{Orbits
  in a non-Kerr Dynamical System}},\ }\href
  {https://doi.org/10.1142/S0218127411029768} {\bibfield  {journal} {\bibinfo
  {journal} {Int. J. Bifurc. Chaos}\ }\textbf {\bibinfo {volume} {21}},\
  \bibinfo {pages} {2261} (\bibinfo {year} {2011})},\ \Eprint
  {https://arxiv.org/abs/1108.5057} {arXiv:1108.5057 [gr-qc]} \BibitemShut
  {NoStop}%
\bibitem [{\citenamefont {Arnol'd}(1963)}]{Arnold_1963}%
  \BibitemOpen
  \bibfield  {author} {\bibinfo {author} {\bibfnamefont {V.~I.}\ \bibnamefont
  {Arnol'd}},\ }\bibfield  {title} {\bibinfo {title} {Proof of a theorem of
  a. n. kolmogorov on the invariance of quasi-periodic motions under small
  perturbations of the hamiltonian},\ }\href
  {https://doi.org/10.1070/RM1963v018n05ABEH004130} {\bibfield  {journal}
  {\bibinfo  {journal} {Russian Mathematical Surveys}\ }\textbf {\bibinfo
  {volume} {18}},\ \bibinfo {pages} {9} (\bibinfo {year} {1963})}\BibitemShut
  {NoStop}%
\bibitem [{\citenamefont {Möser}(1962)}]{Moser:430015}%
  \BibitemOpen
  \bibfield  {author} {\bibinfo {author} {\bibfnamefont {J.}~\bibnamefont
  {Möser}},\ }\bibfield  {title} {\bibinfo {title} {{On invariant curves of
  area-preserving mappings of an annulus}},\ }\href
  {https://cds.cern.ch/record/430015} {\bibfield  {journal} {\bibinfo
  {journal} {Nachr. Akad. Wiss. Göttingen, II}\ ,\ \bibinfo {pages} {1}}
  (\bibinfo {year} {1962})}\BibitemShut {NoStop}%
\bibitem [{\citenamefont {Birkhoff}(1913)}]{Birkhoff:1913}%
  \BibitemOpen
  \bibfield  {author} {\bibinfo {author} {\bibfnamefont {G.~D.}\ \bibnamefont
  {Birkhoff}},\ }\bibfield  {title} {\bibinfo {title} {Proof of poincaré's
  geometric theorem},\ }\href {http://www.jstor.org/stable/1988766} {\bibfield
  {journal} {\bibinfo  {journal} {Transactions of the American Mathematical
  Society}\ }\textbf {\bibinfo {volume} {14}},\ \bibinfo {pages} {14} (\bibinfo
  {year} {1913})}\BibitemShut {NoStop}%
\bibitem [{\citenamefont {Lukes-Gerakopoulos}\ and\ \citenamefont
  {Contopoulos}(2013)}]{Lukes-Gerakopoulos:2013gwa}%
  \BibitemOpen
  \bibfield  {author} {\bibinfo {author} {\bibfnamefont {G.}~\bibnamefont
  {Lukes-Gerakopoulos}}\ and\ \bibinfo {author} {\bibfnamefont
  {G.}~\bibnamefont {Contopoulos}},\ }\bibfield  {title} {\bibinfo {title}
  {{Mind the Resonances: Final stages of accretion into bumpy black holes}},\
  }\href {https://doi.org/10.1088/1742-6596/453/1/012005} {\bibfield  {journal}
  {\bibinfo  {journal} {J. Phys. Conf. Ser.}\ }\textbf {\bibinfo {volume}
  {453}},\ \bibinfo {pages} {012005} (\bibinfo {year} {2013})},\ \Eprint
  {https://arxiv.org/abs/1304.7612} {arXiv:1304.7612 [gr-qc]} \BibitemShut
  {NoStop}%
\bibitem [{\citenamefont {Boshkayev}\ \emph {et~al.}(2014)\citenamefont
  {Boshkayev}, \citenamefont {Bini}, \citenamefont {Rueda}, \citenamefont
  {Geralico}, \citenamefont {Muccino},\ and\ \citenamefont
  {Siutsou}}]{Boshkayev:2014mua}%
  \BibitemOpen
  \bibfield  {author} {\bibinfo {author} {\bibfnamefont {K.}~\bibnamefont
  {Boshkayev}}, \bibinfo {author} {\bibfnamefont {D.}~\bibnamefont {Bini}},
  \bibinfo {author} {\bibfnamefont {J.}~\bibnamefont {Rueda}}, \bibinfo
  {author} {\bibfnamefont {A.}~\bibnamefont {Geralico}}, \bibinfo {author}
  {\bibfnamefont {M.}~\bibnamefont {Muccino}},\ and\ \bibinfo {author}
  {\bibfnamefont {I.}~\bibnamefont {Siutsou}},\ }\bibfield  {title} {\bibinfo
  {title} {{What can we extract from quasi-periodic oscillations?}},\ }\href
  {https://doi.org/10.1134/S0202289314040033} {\bibfield  {journal} {\bibinfo
  {journal} {Grav. Cosmol.}\ }\textbf {\bibinfo {volume} {20}},\ \bibinfo
  {pages} {233} (\bibinfo {year} {2014})},\ \Eprint
  {https://arxiv.org/abs/1412.8214} {arXiv:1412.8214 [astro-ph.HE]}
  \BibitemShut {NoStop}%
\bibitem [{\citenamefont {{Boshkayev}}\ \emph {et~al.}(2015)\citenamefont
  {{Boshkayev}}, \citenamefont {{Rueda}},\ and\ \citenamefont
  {{Muccino}}}]{Boshkayev:2015ARep...59..441B}%
  \BibitemOpen
  \bibfield  {author} {\bibinfo {author} {\bibfnamefont {K.}~\bibnamefont
  {{Boshkayev}}}, \bibinfo {author} {\bibfnamefont {J.}~\bibnamefont
  {{Rueda}}},\ and\ \bibinfo {author} {\bibfnamefont {M.}~\bibnamefont
  {{Muccino}}},\ }\bibfield  {title} {\bibinfo {title} {{Extracting multipole
  moments of neutron stars from quasi-periodic oscillations in low mass X-ray
  binaries}},\ }\href {https://doi.org/10.1134/S1063772915060050} {\bibfield
  {journal} {\bibinfo  {journal} {Astronomy Reports}\ }\textbf {\bibinfo
  {volume} {59}},\ \bibinfo {pages} {441} (\bibinfo {year} {2015})}\BibitemShut
  {NoStop}%
\bibitem [{\citenamefont {Boshkayev}\ \emph {et~al.}(2017)\citenamefont
  {Boshkayev}, \citenamefont {Rueda},\ and\ \citenamefont
  {Muccino}}]{Boshkayev:2016epd}%
  \BibitemOpen
  \bibfield  {author} {\bibinfo {author} {\bibfnamefont {K.}~\bibnamefont
  {Boshkayev}}, \bibinfo {author} {\bibfnamefont {J.~A.}\ \bibnamefont
  {Rueda}},\ and\ \bibinfo {author} {\bibfnamefont {M.}~\bibnamefont
  {Muccino}},\ }\bibfield  {title} {\bibinfo {title} {{Main parameters of
  neutron stars from quasi-periodic oscillations in low mass X-ray binaries}},\
  }in\ \href {https://doi.org/10.1142/9789813226609_0442} {\emph {\bibinfo
  {booktitle} {{14th Marcel Grossmann Meeting on Recent Developments in
  Theoretical and Experimental General Relativity, Astrophysics, and
  Relativistic Field Theories}}}},\ Vol.~\bibinfo {volume} {4}\ (\bibinfo
  {year} {2017})\ pp.\ \bibinfo {pages} {3433--3440},\ \Eprint
  {https://arxiv.org/abs/1604.02398} {arXiv:1604.02398 [astro-ph.HE]}
  \BibitemShut {NoStop}%
\end{thebibliography}%

\end{document}